%% file: main.tex
\documentclass[twocolumn]{bytedance_seed}

\usepackage[toc,page,header]{appendix}

\usepackage{minitoc}

\usepackage{tikz}
\usepackage{amsmath}

\usepackage{xspace}
\usepackage{xcolor}
\usepackage{listings}
\usepackage{enumitem}
\usepackage{multirow}
\usepackage{makecell}
\usepackage{subcaption}
\usepackage{tikz}
\usepackage{algorithm}
\usepackage{algorithmic}
\usepackage[dvipsnames]{xcolor}
\usepackage{adjustbox}

\DeclareRobustCommand{\icon}{%
  \begingroup\normalfont
  \raisebox{-0.1em}{%
  \includegraphics[height=1.2em]{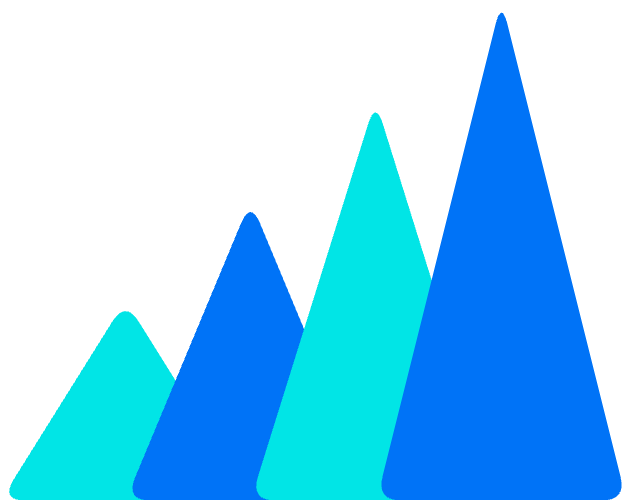}%
  }%
  \kern 0.4em%
  \endgroup
}

\newcommand{\vescale}{\texttt{veScale}\xspace}
\newcommand{\megatron}{Megatron-LM\xspace}
\newcommand{\titan}{TorchTitan\xspace}
\newcommand{\deepspeed}{Megatron-DeepSpeed\xspace}
\newcommand{\rawdeepspeed}{DeepSpeed\xspace}
\newcommand{\colossal}{Colossal-AI\xspace}
\newcommand{\slapo}{Slapo\xspace}
\newcommand{\jax}{JAX\xspace}
\newcommand{\llama}{LLaMA\xspace}
\newcommand{\llamathree}{LLaMA-3\xspace}
\newcommand{\mixtral}{Mixtral\xspace}
\newcommand{\dit}{DiT\xspace}

\newcommand{\bad}[1]{\textcolor{BrickRed}{\textbf{#1}}}
\newcommand{\good}[1]{\textcolor{Green}{\textbf{#1}}}

\definecolor{codecomment}{RGB}{0,128,0}
\definecolor{codenumber}{RGB}{9,134,88}
\definecolor{codestring}{RGB}{163,21,21}
\definecolor{codekey}{RGB}{110,65,198}

\newcommand*\circled[1]{\textcircled{\raisebox{-0.7pt}{#1}}}

\lstdefinestyle{mystyle}{
    commentstyle=\color{codecomment},
    keywordstyle=\ttfamily\footnotesize,
    numberstyle=\tiny\color{codenumber},
    stringstyle=\color{codestring},
    basicstyle=\ttfamily\footnotesize,
    breakatwhitespace=false,         
    breaklines=false,                 
    keepspaces=true,                 
    showspaces=false,                
    showstringspaces=false,
    showtabs=false,                  
    tabsize=2,
    columns=fullflexible,
}
\newcommand{\ic}[1]{\lstinline[basicstyle=\ttfamily\footnotesize]|#1|}

\newcommand{\para}[1]{{\bf \noindent #1 \hspace{1pt}}}

\usepackage{url}
\usepackage{hyperref} % Load hyperref (includes url package)
\hypersetup{
    colorlinks=true,  % Use colors, not boxes
}
\title{\vescale: Consistent and Efficient Tensor Programming with Eager-Mode SPMD}

\author[*,\mathsection]{Youjie Li}
\author[*,\dagger]{Cheng Wan}
\author[]{Zhiqi Lin}
\author[]{Hongyu Zhu}
\author[]{Jiacheng Yang}
\author[\dagger]{Ziang Song}
\author[]{Xinyi Di}
\author[]{Jiawei Wu}
\author[]{Huiyao Shu}
\author[]{Wenlei Bao}
\author[\mathsection]{Yanghua Peng}
\author[]{Haibin Lin}
\author[]{Li-Wen Chang}

\affiliation[]{ByteDance Seed}

\contribution[*]{Co-Primary Author}
\contribution[\dagger]{Work done at ByteDance Seed}
\contribution[\mathsection]{Corresponding authors}

\abstract{
Large Language Models (LLMs) have scaled rapidly in size and complexity, requiring increasingly intricate parallelism for distributed training, such as 3D parallelism. 
This sophistication motivates a shift toward simpler, more debuggable programming paradigm like Single Program Multiple Data (SPMD). 
However, SPMD in eager execution introduces two key challenges: ensuring consistency with single-device execution and achieving high performance at scale.

In this paper, we introduce \vescale, an eager-mode training system that fully embraces SPMD paradigm to democratize distributed tensor programming. 
\vescale addresses the prevalent issue of inconsistent results in systems like PyTorch by introducing a novel algorithm of distributed Random Number Generation (RNG) compatible with arbitrary sharded operators. 
\vescale also significantly boosts training performance by reducing PyTorch primitive's overhead and improving communication efficiency. 
Evaluations show that \vescale delivers up to $2.2\times$ speedup over the state-of-the-art training systems, like \titan, and cuts code complexity by $78.4\%$, while preserving single-device-equivalent results.
}

\checkdata[Project Page]{\url{https://github.com/volcengine/veScale}}

\begin{document}
\maketitle

\input{section/1-Introduction}
\input{section/2-Background}

\input{section/3-Overview}
\input{section/4-API}
\input{section/5-Correctness}
\input{section/6-Performance}
\input{section/7-Implementation}

\input{section/8-Experiments}
\input{section/9-Related_Work}
\input{section/10-Conclusion}
\input{section/11-Acknowledgements}

\clearpage

\bibliographystyle{plainnat}
\bibliography{reference}

\clearpage

\beginappendix

\input{section/appendix}

\end{document}

%% file: section/1-Introduction.tex
\section{Introduction}
\label{sec:intro}

The rapid expansion of Large Language Models (LLMs) has brought unprecedented revolutions in numerous areas such as natural language processing and content generation~\cite{nijkamp2022codegen,kasneci2023chatgpt,li2023chatdoctor,wang2024visionllm}. 
The ever-growing LLMs have led to massive computation and memory demands for training, making distributed training an essential solution to meet these requirements.

Distributed training systems are thus driven to evolve with increasingly complex parallelisms, from a simple data parallel~\cite{li2020pytorch} to the sophisticated N-Dim parallel~\cite{shoeybi2019megatron}.
Although N-Dim parallelism has demonstrated great performance, its programming paradigm has rarely been designed carefully to the extent that model definitions are arbitrarily coupled with distributed system intricacies.
This issue has reached a point where 
model developers cannot focus on designing new model structures without having to manage the intricate parallelization across devices. 
Meanwhile, they cannot enhance the system performance without hacking new model structures.
This has motivated a new trend in programming paradigms that decouple parallelization from model definitions, enabling independent innovations on model architectures and distributed training systems.

Decoupled parallelization from model has been explored in prior works~\cite{lepikhin2020gshard, agrawal2019tensorflow, nnscaler, li2023colossal, xu2021gspmd}. 
Compiler-based systems like Alpa~\cite{zheng2022alpa} and JAX~\cite{jax2018github} allow users to write programs in a Single-Program Multi-Data (SPMD) style, automatically parallelizing model graphs within the compiler's passes. 
However, these systems are often challenging to use and debug~\cite{jax2024traceerror, tf2024debug} compared to eager-based systems that execute model code line-by-line in a script without compilation and offer flexible usage and interactive debug.
Thus, the eager systems like PyTorch~\cite{paszke2019pytorch} dominate the community and own 92\% models on HuggingFace~\cite{assembly2023torchpercent}.
Among the eager ecosystem, the widely adopted systems of training LLMs are \megatron~\cite{shoeybi2019megatron} and \deepspeed~\cite{rasley2020deepspeed}, due to their notable efficiency and scalability.
Nonetheless, they are also known for the heavily intertwined parallelisms with model definitions. 
Fortunately, PyTorch now provides the \textit{D}istributed\textit{Tensor} (\textit{DTensor}) primitive~\cite{torch-dtensor} to enable decoupled parallelization in eager execution. 
DTensor supports flexible parallelization by transparently distributing operators from single-device model definitions, and empowers the state-of-the-art training system, \titan~\cite{torchtitan2024}.
Despite these advancements of eager-mode SPMD, our analysis~(\S\ref{subsec:motivations}) identifies two critical issues that remain unaddressed.

First, PyTorch DTensor faces a consistency issue in maintaining single-device semantics, 
as its parallelization mechanisms do not account for single-device-equivalent randomness~\cite{dtensor2024rngissue}. 
This leads to mismatched operations and weight initialization between single-device and distributed execution (as well as between different parallelizations),
resulting in inconsistent training outcomes, which plague reproducibility and debuggability. 
Second, DTensor encounters performance issues due to its eager execution, where every operator needs to go through a complex dispatch logic at runtime, leading to significant CPU overhead (e.g., 58\% slower than vanilla Tensor~\cite{dtensor2023cpuissue1}) and costly communication~\cite{dtensor2023cpuissue2}. 

To address all these challenges, this paper presents \vescale, an eager-mode distributed training system that fully adopts the SPMD programming, ensuring both \textit{consistent single-device semantics} and \textit{high-performance runtime} with a \textit{simple interface}.
Our key contributions are as follows:

\noindent $\triangleright$ \vescale requires \textit{zero change on model definition} for parallelization by offering simple APIs to fully decouple distributed intricacies.
This empowers users to build models as if on a single device, and lets \vescale parallelize models transparently while preserving the power of eager for interactive execution and debug in distributed environment.
Our statistics show that \vescale saves $\sim$78.4\% development effort~(\S\ref{sec:api}).

\noindent $\triangleright$ \vescale guarantees consistent single-device semantics by introducing a novel distributed Random Number Generation (RNG) algorithm. 
It ensures bit-wise match of arbitrary random tensors between single-device and distributed RNG, regardless of different placements or device counts.
This enables users to compare/reproduce/debug training outcomes reliably, even under randomness throughout distributed training, such as weight initialization and dropout.
\textit{To the best of our knowledge, such single-device semantic distributed RNG is the first of its kind in the community}~(\S\ref{sec:correctness}).

\noindent $\triangleright$ \vescale identifies the prohibitive CPU overhead incurred by DTensor and eliminates entire overhead through a novel \textit{Static Eager} execution mode as well as three highly effective optimizations.
Additionally, \vescale improves DTensor communication with fusion across multiple \textit{parallel dimensions} and multiple DTensors into a single operation.
With all optimizations, \vescale speeds up DTensor by $\sim$5.21$\times$~(\S\ref{sec:perf_optimization}).

\noindent $\triangleright$ \vescale is evaluated with diverse model structures of different sizes, including dense (\llama~\cite{dubey2024llama}), sparse (\mixtral~\cite{jiang2024mixtral}), and vision models (\dit~\cite{peebles2023scalable}).
\vescale achieves \textit{$\sim2.2\times$ end-to-end speedup} compared to the state-of-the-art systems like \titan, \megatron, and \rawdeepspeed, while offering consistent single-device results and requiring minimal development efforts~(\S\ref{sec:evaluation}).

\vescale has been deployed for internal production 
for numerous tasks ranging from pre-training to fine-tuning of LLMs. 
\textit{We will release an open source version soon}.

%% file: section/2-Background.tex
\section{Motivation}
\label{sec:background}

\subsection{Programming Dilemma of Distributed Training}
\label{subsec:background:training_parallellism}

Fitting modern LLMs on a single device for training becomes infeasible, driving the need for distributed training.
To enable system efficiency, various parallelisms have been developed, including data parallel (DP)~\cite{li2020pytorch}, tensor parallel (TP)~\cite{shoeybi2019megatron}, sequence parallel (SP)~\cite{korthikanti2023reducing}, and optimizer parallel (ZeRO)~\cite{rajbhandari2020zero}.

To support these parallelisms but without coupling the model definition, compiler experts developed several distributed programming paradigms~\cite{lepikhin2020gshard, agrawal2019tensorflow, nnscaler, xu2021gspmd, li2023colossal}. 
Especially, Alpa~\cite{zheng2022alpa} and JAX~\cite{jax2018github} allow users to write SPMD programs and parallelize models based on static model graphs.
However, they often depend on successfully tracing the entire model graph, which is hard to guarantee in practice where non-traceable code are common, such as Python objects, communications, and control flows~\cite{jax2024traceerror,tf2024knowissue}.
Also, compiled programs can be hard to debug~\cite{tf2024debug, jax2024debug}, let alone distributed ones.

By contrast, eager-based systems does not rely on tracing nor compilation, offering flexible programming and interactive debugging.
Thus, eager systems like PyTorch dominate 92\% models on HuggingFace~\cite{assembly2023torchpercent} and 70\% research on PapersWithCode~\cite{paperwithcode2024}.
The most adopted eager training system for LLMs are the \megatron~\cite{shoeybi2019megatron, mdeepspeed2022github} family.
However, they are designed for performance instead of easy programming, and require intrusively rewriting the entire model with intertwined distributed logic, heavily burdening the model developers.
For instance, to support TP, users are required to carefully rewrite original model definition's \ic{Linear}s with \ic{RowParallelLinear} and \ic{ColumnParallelLinear} under multi-device semantics. 
Similar efforts also apply to refine \ic{LayerNorm} for SP and rewrite \ic{Adam} for ZeRO.
It is also users' responsibility to manually invoke communication on \ic{LayerNorm}s' gradient for SP, besides for DP; otherwise, gradients are incorrect.

\subsection{The Promising DTensor}
\label{subsec:background-dtensor-workflow}

\begin{figure}[t]
    \centering
    \includegraphics[width=\linewidth]{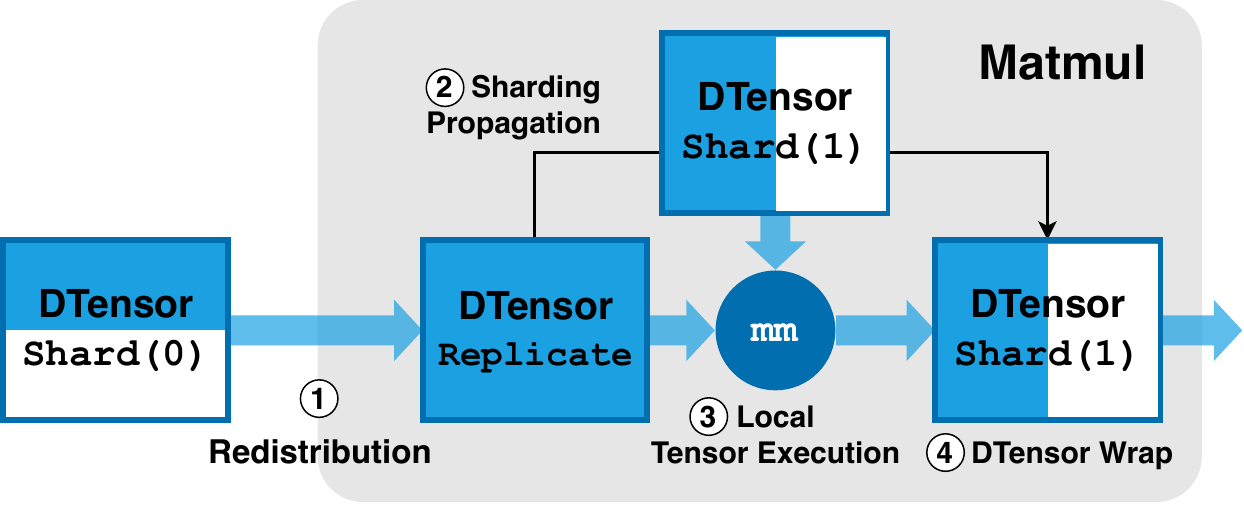}
    \vspace{-1em}
    \caption{DTensor workflow of executing a sharded matrix multiplication (\ic{mm}) on a device. The shaded part in each DTensor indicates the materialized local tensor on this device.
    Note for terminology that \circled{1} is equal to \textit{DTensor Communication}, and \circled{2}\circled{3}\circled{4} compose \textit{DTensor Dispatch}.}
    \label{fig:dtensor}
    \vspace{-1em}
\end{figure}

PyTorch now offers a promising primitive named DTensor~\cite{torch-dtensor} that provides the opportunity towards decoupled parallelization in eager execution.
It represents a global tensor distributed across a set of devices (referred to as a \textit{device mesh}), where each device in the mesh holds a \textit{local tensor} that represents a shard of the global tensor.
DTensor supports three types of \textit{placement} for sharding a global tensor: \textbf{\ic{Shard(dim)}}, \textbf{\ic{Replicate}}, and \textbf{\ic{Partial}}. 
It also enables users to switch between these placement via \textit{redistribution}, which is achieved by implicitly triggering collective communications.
Additionally, DTensors can be computed directly by operators, where each operator owns valid sharding strategies that describe expected input and output placement.

Figure~\ref{fig:dtensor} shows the workflow of DTensor. 
\circled{1} First, the input DTensor is \textit{redistributed} from its source placement to the expected placement for a valid sharded operation, e.g., column-wise matrix multiply.
\circled{2} Next, the sharding metadata of DTensors (placement, global shape, global stride, etc.) is \textit{propagated} from input to output by performing an metadata inference on the CPU, generating the output's metadata.
\circled{3} Then the local tensors are extracted from input DTensors and \textit{executed} by the operator to produce local output on each GPU device. 
\circled{4} Finally, each local output tensor, along with its global output metadata, is \textit{wrapped} back into a DTensor. 

\input{table/movt_rng_diff}

\subsection{Challenges of DTensor}
\label{subsec:motivations}

Despite DTensor's success in enabling parallelization, it faces critical challenges in consistent semantics and performance.

\para{Poor Consistency.}
Ensuring consistent semantics during parallelization is essential for large-scale distributed training and debugging.
Following the single-device programming paradigm, we argue that the consistent semantics of DTensor is \textit{single-device semantics}, meaning that \textit{merging all local tensors across devices should yield the exact same global tensor on a single device}.
For example, local tensors are $[1, 2]$ on GPU 0 and $[3, 4]$ on GPU 1, which can be merged to match the global tensor $[1, 2, 3, 4]$ on a single GPU.
However, the single-device semantics can be easily violated due to undesirable RNG designs or completely missing the fundamental for distributed RNG, e.g., random tensors are generated as $[1, 2]$ on GPU 0 and $[5, 6]$ on GPU 1, which does not match the global random tensor $[1, 2, 3, 4]$ on a single GPU.
We observed that violating this semantics can produce severe consequences, as minor numerical inconsistencies can be amplified across varying device counts, potentially degrading training accuracy or, in the worst case, causing model divergence~\cite{randomness-1,randomness-2}.
Such inconsistencies are extremely difficult to diagnose in production environments, substantially complicating both debugging and reproducing distributed training at scale.

\begin{figure}[!t]
\centering    
    \begin{subfigure}[b]{0.48\linewidth}
         \centering
         \includegraphics[width=\linewidth]{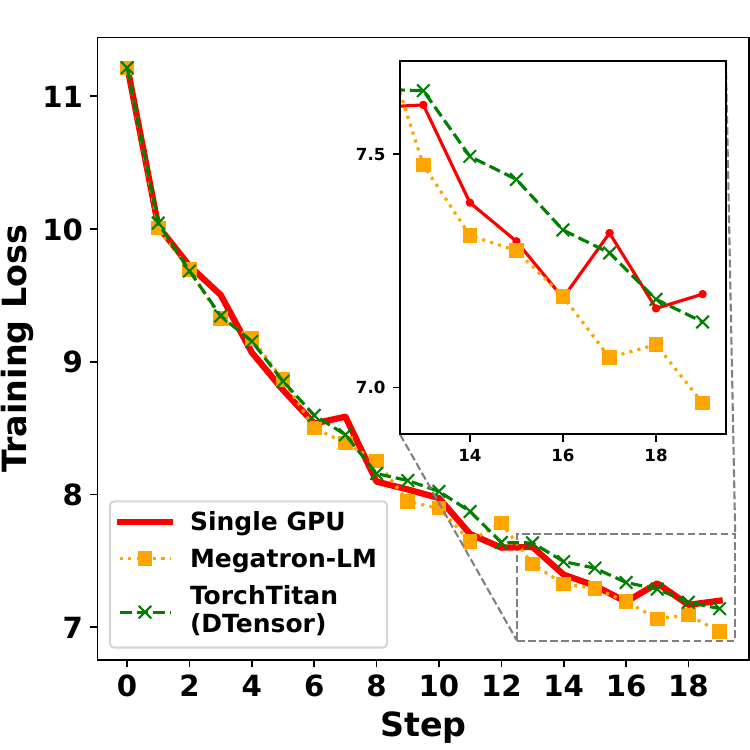}
         \vspace{-1.8em}
         \caption{
         Loss comparison between eight devices with single device.
         }
        \label{fig:related_work_incorrectness}
    \end{subfigure}
    \hfill
    \begin{subfigure}[b]{0.48\linewidth}
         \centering
         \includegraphics[width=\linewidth]{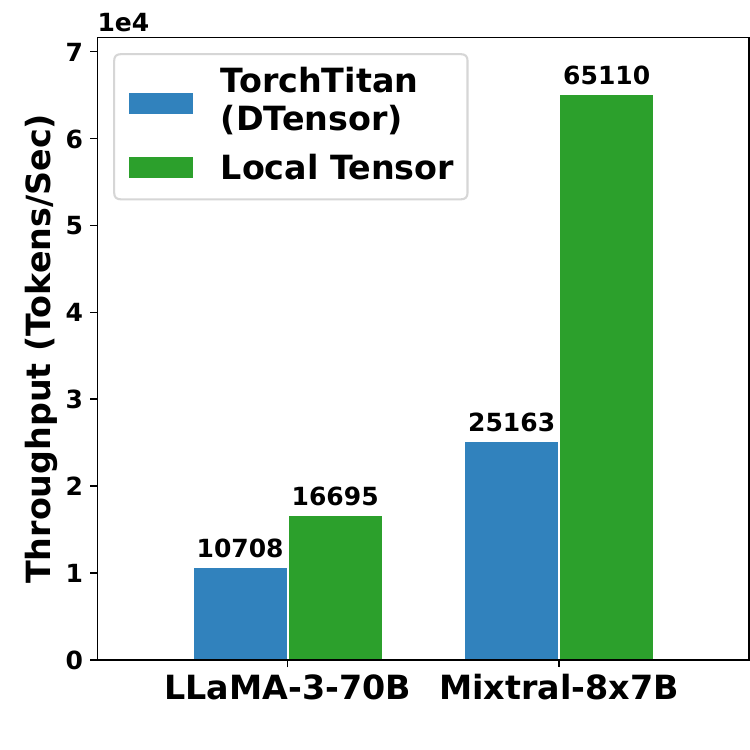}
         \vspace{-1.8em}
         \caption{DTensor overhead in end-to-end throughput on 32 devices.
         }
         \label{fig:related_work_performance}
    \end{subfigure}
    \vspace{-0.5em}
    \caption{Training loss and performance using DTensor.}
    \label{fig:misaligned_loss_and_low_performance}
    \vspace{-1em}
\end{figure}

Unfortunately, the current DTensor does not adhere to single-device semantics, which is even deemed \textit{impossible} to achieve~\cite{dtensor2024rngissue}.
Table~\ref{tab:rng_diff} compares weight initialization using DTensor and single device, where large difference is observed in \textit{four orders of magnitudes}.
Figure~\ref{fig:related_work_incorrectness} further compares training loss curves of \llama 7B, using correct weight initialization, the same random states, and the same hyper-parameters; however, a non-trivial difference is still observed.
This mismatch is not only observed in DTensor but also recognized in most systems such as \megatron (shown in Table~\ref{tab:rng_diff} and Figure~\ref{fig:related_work_incorrectness}) and \rawdeepspeed (e.g., an open issue~\cite{deepspeed2024issue}).

\para{Poor Performance.} DTensor also suffers from poor performance. 
Figure~\ref{fig:related_work_performance} compares end-to-end training performance between \titan with DTensors and local-tensor training without DTensor.
We observe that DTensor is significantly slower than local tensors: 35.8\% for \llama and 61.3\% for \mixtral.
Although \titan already attempted to improve by using local-tensor execution for selected model parts, it still suffers from bottlenecks on remaining DTensors~\cite{dtensor2023cpuissue1,dtensor2023cpuissue2}.

We further identify that the poor performance of DTensor is due to two factors: 1) high CPU overhead during dispatch and 2) inefficient communication.
The high CPU overhead arises as every operator needs to go through a complex dispatch logic at runtime (recall Figure~\ref{fig:dtensor}), often costing over $0.5$ ms of CPU time, even for fast operators like \ic{Tensor.view}.
Given a model with thousands of operators, the total overhead cannot be hidden by the asynchronous GPU execution, leading to GPU idle. 
Also, the DTensor communication accounts for $\sim$11.3\% overhead. 
Such inefficiency comes from fine-grained communication at per-tensor granularity and per-parallel dimension, failing to fully utilize the bandwidth.

%% file: table/movt_rng_diff.tex
\begin{table}[t]
\caption{Mismatched model weights between single-device and distributed initialization on eight devices, shown in terms of element-wise maximal and mean difference. Weights are initialized with $\mathcal{N}(0, 0.025)$.}
\vspace{-0.5em}
\label{tab:rng_diff}
\centering
\footnotesize
\begin{tabular}{| c | cc | cc |}
\hline
 & \multicolumn{2}{c|}{\titan (DTensor)} & \multicolumn{2}{c|}{\megatron}  \\
 & Max & Mean & Max & Mean \\
\hline
Llama-3 8B & {\bf 0.1819} & 0.0197 & {\bf 0.1632} & 0.0197 \\
Mixtral 8x7B & {\bf 0.1948} & 0.0197 & {\bf 0.1818} & 0.0197 \\
\hline
\end{tabular}
\vspace{-1em}
\end{table}

%% file: section/3-Overview.tex
\section{\vescale Overview}
\label{sec:overview}

\begin{figure}[!t]
    \centering
    \includegraphics[width=\linewidth]{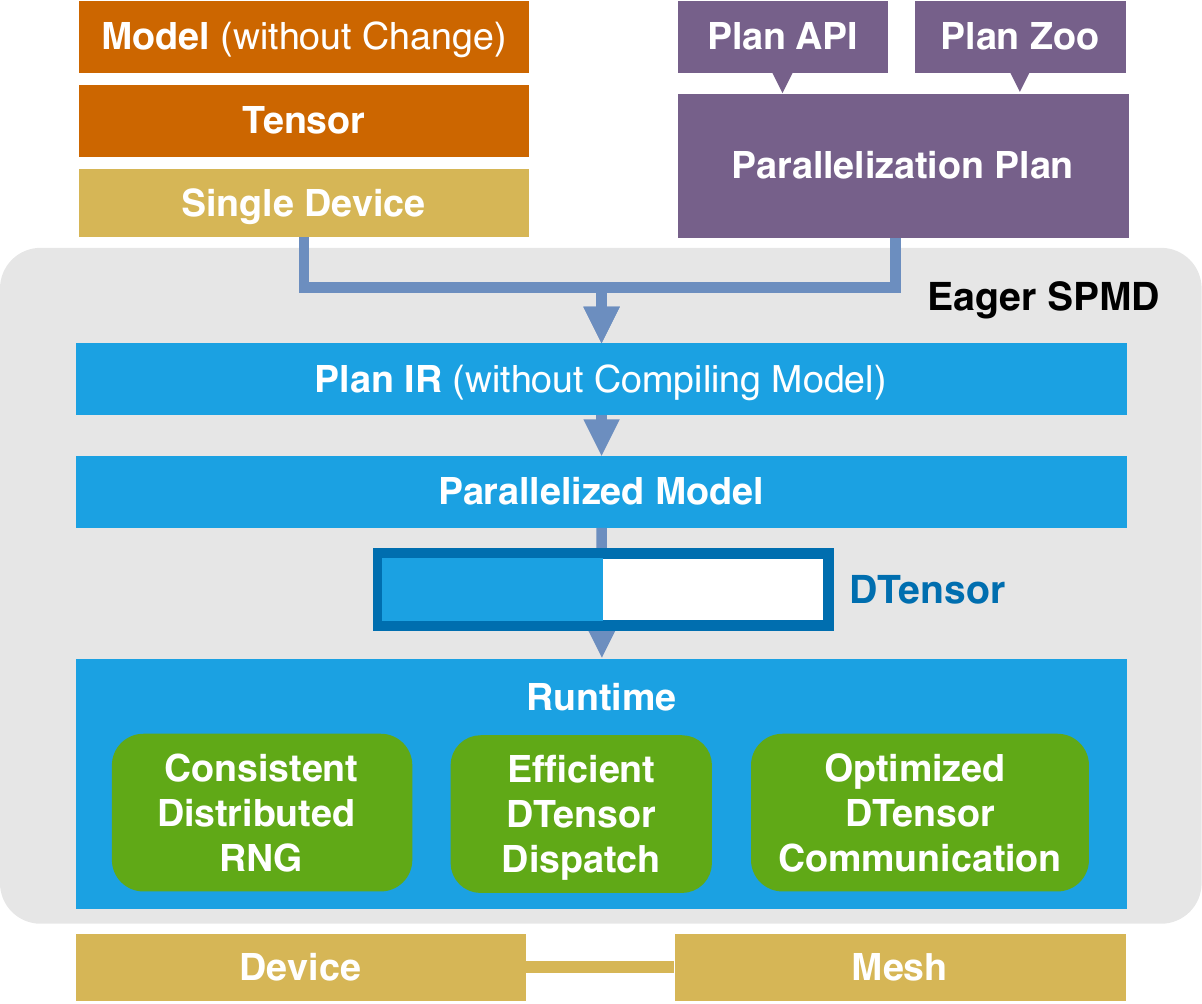}
    \vspace{-1em}
    \caption{\vescale overview.}
    \label{fig:overview}
    \vspace{-1em}
\end{figure}

To address these challenges, we introduce \vescale, an eager SPMD system developed on DTensor, delivering consistent semantics and high performance with easy-to-use interfaces.

Figure~\ref{fig:overview} provides the overview. 
To parallelize model for distributed training, \vescale takes two inputs: 
i) a single-device model definition of PyTorch without requiring any changes or coupling any parallelism,
and ii) a parallelization Plan that describes parallel strategies, such as how to shard weights (\S\ref{sec:api}).
Note that the Plan is orthogonal to the model and can be reused across different models/layers via a Plan Zoo. 
Next, the Plan is transformed into the internal Plan IR (Intermediate Representation) without requiring compilation/trace of the model graph, and is further applied onto the model for parallelization.
This process involves replacing weight Tensors with DTensors and generating PyTorch hooks to callback DTensor redistribution at runtime to ensure proper sharding specified by the Plan.
At runtime, the training script runs DTensors in eager execution with: 
i) a consistent distributed RNG to ensure single-device semantics for all randomness (\S\ref{sec:correctness}),
ii) an efficient DTensor dispatch to minimize its CPU overhead (\S\ref{subsec:dtensor_dispatch_perf_optimization} and \S\ref{subsec:zero_dtensor_dispatch}),
and iii) an optimized DTensor communication to improve bandwidth utilization (\S\ref{subsec:optimize-comm}).

%% file: section/4-API.tex
\section{Easy-to-Use API}
\label{sec:api}

\begin{figure}[t!]
    \centering
    \includegraphics[width=1\linewidth]{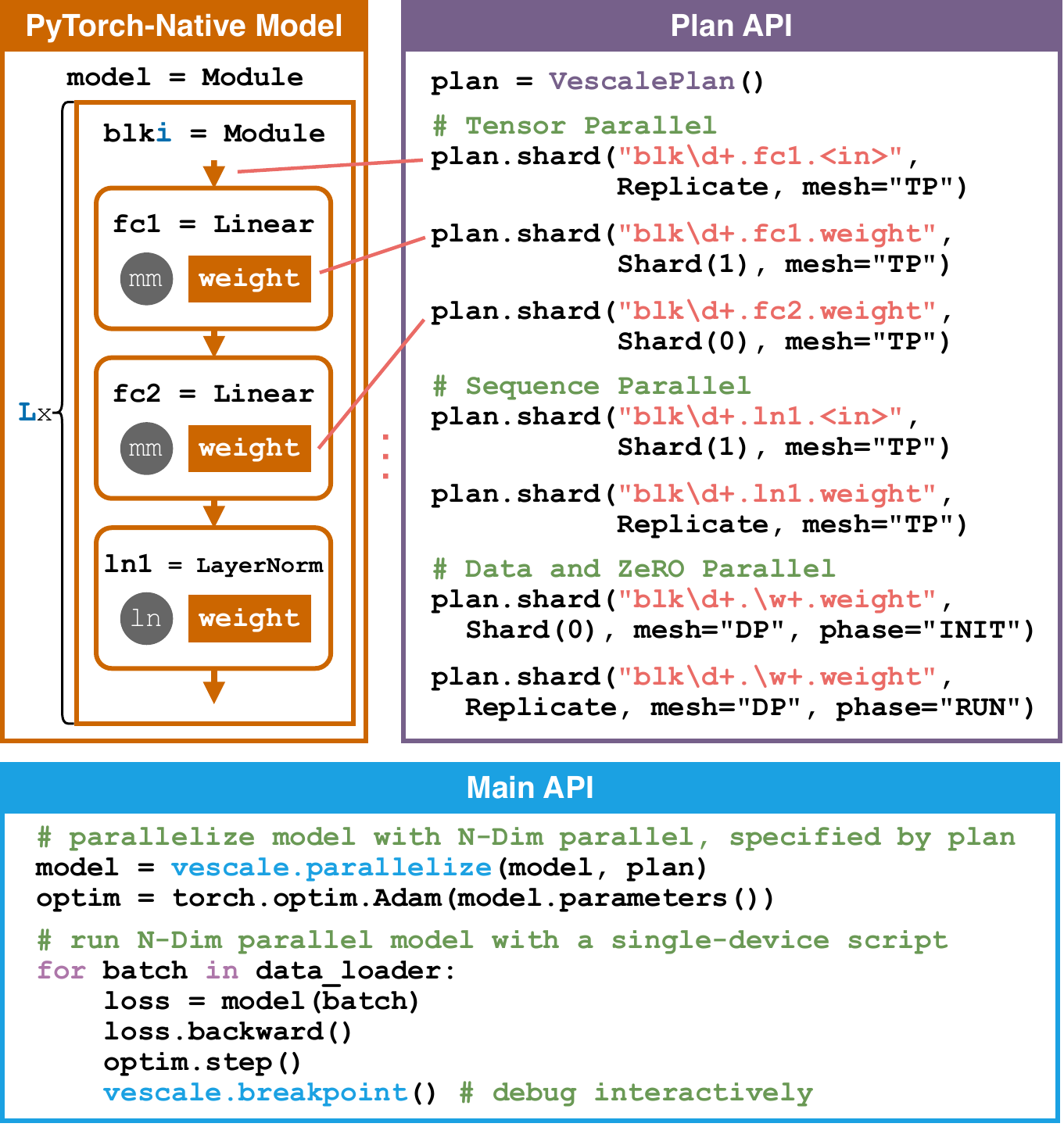}
    \vspace{-1em}
    \caption{\vescale API for eager SPMD.}
    \vspace{-1em}
    \label{fig:api}
\end{figure}

Figure~\ref{fig:api} shows two APIs of \vescale: \textit{Plan API} specifies N-Dim parallel decoupled from a model definition and generates a Plan, and \textit{Main API} applies the Plan for execution.

\para{Plan API.}
The \textbf{\ic{VescalePlan}} API enables flexible parallelisms through descriptive specification of \textit{how} to shard \textit{which} tensor and \textit{when}: \textbf{\ic{plan.shard(tensor\_path, placement, mesh, phase)}}.

As shown in Figure~\ref{fig:api}, this API precisely targets \textit{which} tensor via \ic{tensor\_path}.
This is achieved by leveraging the built-in Module hierarchy of a PyTorch model definition (e.g., nested \ic{nn.Module} and \ic{nn.Linear} with a path \ic{blk1.fc1}) plus the tensor name (e.g., a parameter named \ic{weight} and input activation denoted as \ic{<in>}).
This API also supports regular expressions to match multiple tensors (e.g., \ic{blk\\d+} for all \ic{blk}s), which is essential for ease-of-use when working on repeated model structures like LLMs. 
Upon targeting tensor(s), this API can specify \textit{how} to shard with desired \ic{placement} and \ic{mesh} for flexible parallelism.
E.g., \megatron-style tensor parallel can be specified by sharding \ic{fc1.<in>} to \ic{Replicate}, \ic{fc1.weight} to \ic{Shard(1)}, and \ic{fc2.weight} to \ic{Shard(0)}, with all on a (sub) device mesh named \ic{TP}.
Specifying sequence or data parallel is similar, except that data parallel is often placed on another sub device mesh (e.g., \ic{DP}) or another dimension of the N-Dim device mesh.
Furthermore, rather than fixing the sharding across different time phases of training, this API also allows specifying \textit{when} to shard by separating different placements at different \ic{phase}s, enabling advanced parallels like FSDP~\cite{zhao2023pytorch}/ZeRO~\cite{rajbhandari2020zero}.
E.g., ZeRO-3 can be specified by sharding all weights to \ic{Shard(0)} at initialization (\ic{phase=INIT}) and then resharding to \ic{Replicate} at forward/backward (\ic{phase=RUN}).
Lastly, this API does not demand specifying sharding of gradients, as it is handled automatically under the hood with auto-differentiable hooks.

\begin{figure*}[!t]
    \centering
    \includegraphics[width=\linewidth]{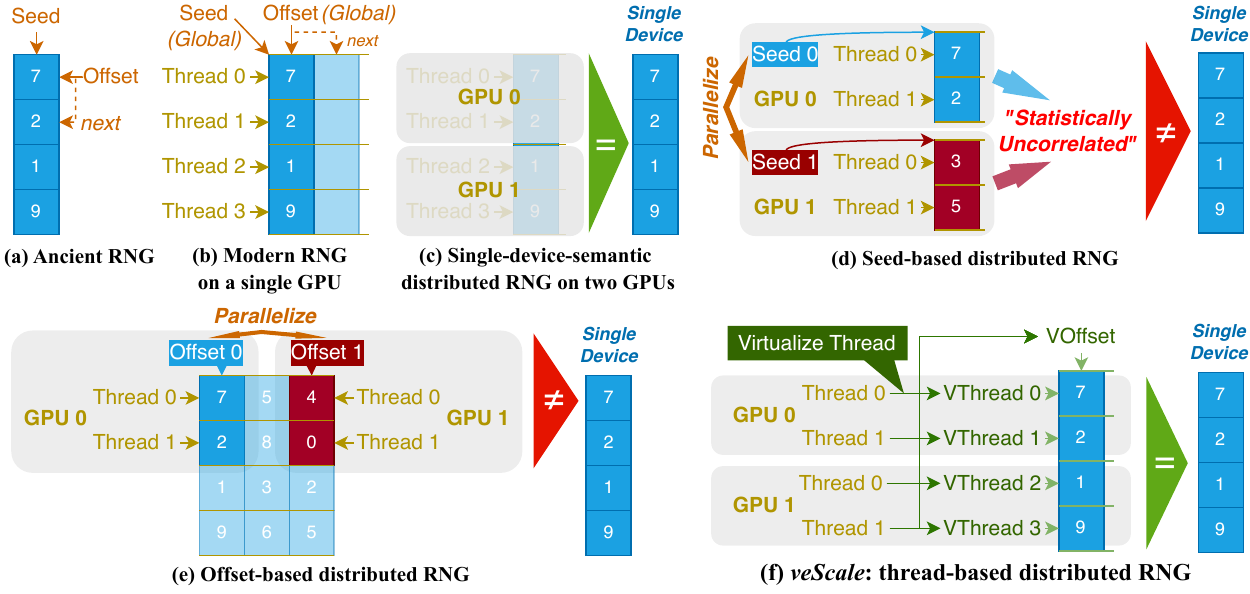}
    \vspace{-1em}
    \caption{Comparison of RNG designs with a simple example of generating random numbers $[7, 2, 1, 9]$. (a) is out-dated but still assumed by many; (b) is used by PyTorch on a single GPU; (c) is the goal to achieve;  (d) is used by most works (\megatron, \rawdeepspeed, and \jax); (e) is used by \titan; (f) is proposed by \vescale for \textit{single-device-semantic distributed RNG}.
    }
    \label{fig:rng_compare}
    \vspace{-1em}
\end{figure*}

\para{Main API.}
The \textbf{\ic{parallelize(model, plan)}} API applies \ic{plan} onto unchanged \ic{model} to initialize an N-Dim parallelized model (recall Figure~\ref{fig:overview}).
In this process, Plan IR takes the key role for transformation and optimization of the Plan logic.
E.g., the first \ic{plan.shard} in Figure~\ref{fig:api} is lowered into an IR abbreviated as \ic{blk1.fc1.forward\_pre:<in>:0:redist(->R,TP)}.
Especially, the IR transformation not only resolves regex path (\ic{blk1}), but also translates \ic{plan.shard} to representation of a correct PyTorch hook (\ic{forward\_pre}) and a desired hook function for DTensor redistribution (\ic{redist}).
Then the IR optimization \textit{fuses multiple IRs} with the same path and hook into one, and further batches their hook functions with \textit{duplication} and \textit{valid reordering}.
Afterwards, the optimized IRs are converted into real hooks which are installed on the model with callback execution at runtime with minimal overhead.
Meanwhile, the IR transformation also figures out when hooks are not needed, such as the parameter sharding or \ic{INIT} phase, and directly converts them into DTensors.

After \ic{parallelize}, the optimizer is also parallelized automatically by just using a \textit{PyTorch-native optimizer} taking DTensor weights of the model, which again decouples the heavy complexity of optimizer parallel like \megatron's DistributedOptimizer~\cite{megatron2024distoptim}.
Now, with both parallelized model and optimizer, users can enjoy parallel training with only a \textit{single-device script}, and let \vescale transparently handle all runtime complexity (consistency and performance).
Further, empowered by eager execution, users can even \textit{interactively debug} the runtime with arbitrary breakpoints and line-by-line steps.
To achieve this in distributed environment, \vescale extends PDB~\cite{python2025pdb} and offers \textbf{\ic{breakpoint()}} API.

%% file: section/5-Correctness.tex
\section{Consistent Semantics}
\label{sec:correctness}

This section explains how \vescale achieves consistent single-device semantics of SPMD training.

\para{Root Cause.}
Upon investigating the misaligned single-device semantics in existing systems~\cite{shoeybi2019megatron, rasley2020deepspeed, chen2024slapo, torchtitan2024, li2023colossal, deepspeed2024issue, jax2018github}, we identify the root cause as misaligned distributed \textit{R}andom \textit{N}umber \textit{G}eneration (\textit{RNG}). 
Specifically, the distributed RNG result does not match the original RNG result on a single device (\S\ref{subsec:motivations}). Although existing works have attempted simple seeding or offsetting of the RNG state between devices, these solutions are still insufficient to achieve the \textit{single-device-semantic} RNG.
Furthermore, inconsistent distributed RNG impacts the training from end to end, such as in distributed weight initialization, distributed data sampling, and distributed operators like dropout, thus compromising the single-device semantics of the entire training.

Figure~\ref{fig:rng_compare} summarizes all existing RNG designs. 
A common but outdated concept of RNG is shown in {\bf (a)} where a random sequence is generated sequentially from an initial seed and is stepped by element-wise offset.
However, modern designs utilize a GPU for parallel RNG {\bf (b)} where a random sequence is generated in parallel with each thread rendering an element under the same global seed and the same global offset, and the RNG continues with all threads stepping to the next global offset. Hence, each random element is determined by three factors: {\textit{\bf seed}}, {\textit{\bf offset}}, and {\textit{\bf thread index}}. 
When scaling up to distributed RNG {\bf (c)}, the original random sequence is sharded across multiple GPUs, where each GPU generates a subsequence but all GPUs together generate the original sequence on the single GPU, which is the desired result.

Regrettably, most works opt to parallelize the seed {\bf (d)}, assigning each GPU a \textit{different local seed} (typically incremented by each GPU's rank~\cite{megatron2019rng, deepspeed2020rng, colossal2021rng}). 
This leads to subsequences that are \textit{statistically uncorrelated}~\cite{nvidia2024pseudorandom} and mismatch the single-device result.
Other works, such as \titan, parallelize the offset {\bf (e)}, assigning a \textit{different local offset} (usually incremented by local subsequence size~\cite{titan2024rng}) to each GPU for generating its subsequence.
This is equivalent to non-contiguously sample numbers in the original random sequence, thus still mismatching the single-device result. 

In this work, \vescale proposes a \textit{thread-based} distributed RNG design {\bf (f)}, where threads and offsets are virtualized as if operating within a global virtual GPU. Each local thread on a physical GPU is mapped to a virtual thread, generating the desired random subsequence as if it were from the global virtual GPU. Thus, \vescale ensures single-device semantics for distributed RNG and SPMD training.

\begin{figure}[!t]
    \centering
    \includegraphics[width=1\linewidth]{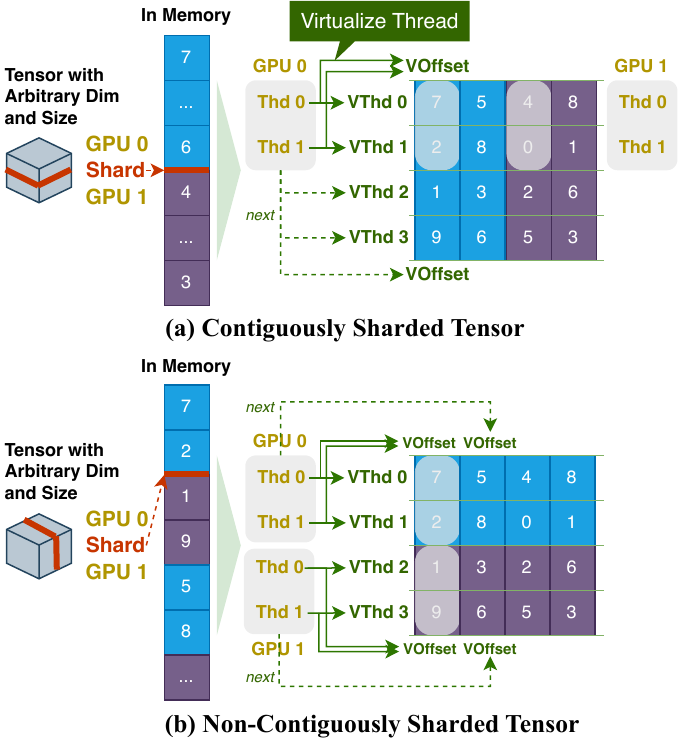}
    \vspace{-1em}
    \caption{\vescale's single-device semantic distributed RNG supporting \textit{arbitrary tensor dimensions, sizes, and both contiguous and non-contiguous sharding}.}
    \label{fig:rng_dive}
    \vspace{-1em}
\end{figure}

\para{Single-Device Semantic Distributed RNG.}
Figure~\ref{fig:rng_dive} illustrates two examples of generating distributed random tensors with single-device semantics.
The flow is as below: 
\begin{enumerate}[leftmargin=*,noitemsep]
  \vspace{-0.25em}
  \item  A distributed random tensor is defined with arbitrary dimension/size and is arbitrarily sharded across GPUs.
  \item The tensor is flatten in memory (global view) and can be contiguous (a) or non-contiguous (b).
  \item Each GPU holds a local tensor that is also flattened in memory and is folded against number of local threads.
  \item Each local thread wants to generate a random value to fill an element in the local tensor at index $i$.
  \item To do so, local index $i$ is first mapped to the global index $j$ in the distributed tensor, and then $j$ is mapped to the global virtual thread and offset.
  \item With virtual thread and offset plus a constant global seed, a backend random value generator (e.g., \ic{curand}~\cite{nvidia2024pseudorandom}) is invoked to generate a random value for the element at $i$.
  \item Each local thread then advances to the next local index by the number of local threads and repeats the step 4.
  \item All local tensors are generated concurrently on their GPUs with possibly different virtual threads and offsets.
  \vspace{-0.25em}
\end{enumerate}
Algorithm~\ref{alg:rng} details the formal flow (with simplification).

\input{algorithm/rng}

%%%%%%%%%%%%%%%%%%%%%%%%%%%%%%%%%%%%%%%%%%%%%%%%%%%%%%%%%%%%%%%%%%%%%%%%%%%%%%%%%
\begin{figure*}[t!]
    \centering
    \includegraphics[width=1\linewidth]{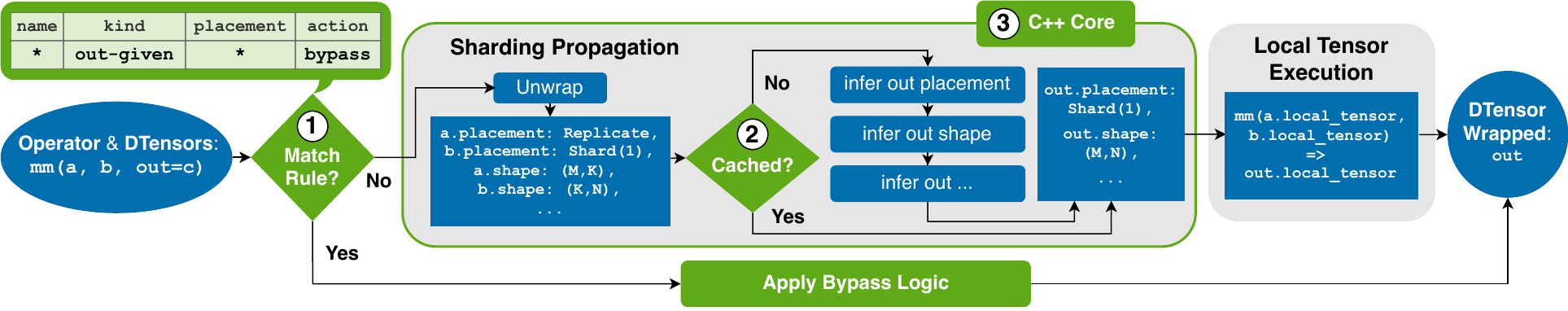}
    \vspace{-1em}
    \caption{DTensor dispatch process (shown in ``blue'') with \vescale's three performance optimizations (shown in ``green''). Illustrated by a simplified example of sharded matrix multiplication (\ic{mm}).} % (with simplification).}
    \vspace{-1em}
    \label{fig:dtensor_dispatch_opt}
\end{figure*}
%%%%%%%%%%%%%%%%%%%%%%%%%%%%%%%%%%%%%%%%%%%%%%%%%%%%%%%%%%%%%%%%%%%%%%%%%%%%%%%%%

%% file: algorithm/rng.tex
\newcommand{\Comment}[1]{\textcolor{blue!50!black}{\footnotesize// #1}}
\newcommand{\IndentNo}{\hspace{5em}}
\newcommand{\IndentIn}{\hspace{3em}}
\newcommand{\floor}[1]{\left\lfloor #1 \right\rfloor}
\newcommand{\ceil}[1]{\left\lceil #1 \right\rceil}
%%%%%%%%%%%%%%%%%%%%%%%%%%%%%%%%%%%%%%%%%%%
\begin{algorithm}[!t]
\footnotesize
\caption{\vescale's Single-device-semantic distributed RNG.}
\label{alg:rng}
\begin{algorithmic}
    \STATE \Comment{A parallel algorithm in per-thread view on local GPU}
    \STATE {\bfseries Input:} local tensor flattened in memory $\chi$,
    \STATE \IndentIn local thread index $\tau$,
    \STATE \IndentIn global random seed $\alpha$ and global offset $\beta$,
    \STATE \IndentIn \Comment{Meta data below; got from DTensor and GPU}
    \STATE \IndentIn local tensor shape $\upsilon$,
    \STATE \IndentIn global tensor shape $\omega$ and stride $\pi$,
    \STATE \IndentIn local tensor coordinate in global tensor $\rho$,
    \STATE \IndentIn local thread count $\theta$ and global thread count $\Theta$,
    \STATE \IndentIn local random value generator $\lambda$ (e.g., \ic{curand})
    \STATE {\bfseries Output:} local tensor filled with desired random values $\chi$ 
    \STATE Get dimension count $D \leftarrow |\upsilon|$
    \STATE Get local element count $I \leftarrow |\chi|$
    \STATE \Comment{Generate random value for local tensor element at $i$}
    \FOR{$i \leftarrow \tau$, $i \leftarrow i + \theta$, while $i < \ceil{I / \theta} \theta$}
        \IF{$i < I$}
            \STATE \Comment{Map local index $i$ to global index $j$ in flattened memory}
            \STATE $i' \leftarrow i$
            \STATE $j \leftarrow 0$
            \FOR{$d \leftarrow D - 1$ {\bfseries to} $0$}
                \STATE Get coordinate $c_d \leftarrow \rho_{d} + i' \bmod \upsilon_d$
                \STATE $j \leftarrow j + c_d \cdot \pi_d$
                \STATE $i' \leftarrow \floor{i' / \upsilon_d}$
            \ENDFOR
            \STATE \Comment{Virtualize thread index and offset}
            \STATE $\tau_{virtual} \leftarrow j \bmod \Theta$ 
            \STATE $\beta_{virtual} \leftarrow \floor{j / \Theta}$ 
            \STATE \Comment{Generate random value $r$ in global view}
            \STATE $\beta_{virtual} \leftarrow \beta_{virtual} + \beta$
            \STATE $r \leftarrow \lambda(\alpha, \tau_{virtual}, \beta_{virtual})$ 
            \STATE $\chi_i \leftarrow r$ 
        \ENDIF
    \ENDFOR
\end{algorithmic}
\end{algorithm}
%%%%%%%%%%%%%%%%%%%%%%%%%%%%%%%%%%%%%%%%%%%

%% file: section/6-Performance.tex
\section{Performance Optimization}
\label{sec:perf_optimization}
This section introduces the primary techniques developed by \vescale to achieve high end-to-end performance.

\subsection{Efficient DTensor Dispatch} 
\label{subsec:dtensor_dispatch_perf_optimization}

\para{The Slow DTensor Dispatch.}
As discussed in \S\ref{subsec:motivations}, DTensor incurs significant CPU overhead compared with local Tensor.
We found the root cause to be the \textit{DTensor Dispatch} process, which involves not just local Tensor execution but also complex computation to infer the metadata of output DTensor(s) (including placement, global shape, global stride, etc.).
Such inferring requires parsing the incoming operator and all its inputs, and then propagates the metadata from input DTensors to output DTensor(s) -- \textit{Sharding Propagation}.
Only with the inferred metadata and local output tensor, the output DTensor can be produced.
Figure~\ref{fig:dtensor_dispatch_opt} shows a simplified DTensor dispatch.

We observe that Sharding Propagation is the major overhead (up to 580 $\mu$s per operator dispatch) and is at least 7$\times$ longer than local tensor execution (i.e., the real work).
This observation is consistent with findings in~\cite{dtensor2023cpuissue1}. 
To be specific, Sharding Propagation is heavily plagued by inferring shape/stride using Fake Tensors~\cite{torch2025faketensor} that contain only metadata without real tensor data. 
Such inferring runs operators on CPU with input Fake Tensors to produce output Fake Tensor for using its shape/stride. 
It is deemed as \textit{pretty slow in practice} by the PyTorch officials~\cite{torch2025slowfaketensor}.

To reduce the unbearable tax of DTensor dispatch, we introduce three effective optimizations, 
as shown in Figure~\ref{fig:dtensor_dispatch_opt}:

\para{\circled{1} Rule-based Bypass.}
It is motivated by the fact that certain operators have their output metadata already known, e.g., 
\ic{mm(a, b, out)} with a given output placeholder containing metadata or \ic{equal} with a boolean output without metadata, which allows us to bypass DTensor dispatch entirely.
Inspired by network rules, we designed \textit{DTensor Dispatch Rules} to match and act as bypassing the dispatch at entry or halfway.
Incoming operators and input DTensors are matched against our rules that are defined by four fields: \textbf{\ic{<op-name, op-kind, input-placement, custom-meta>}}.
E.g., a rule: 
\begin{itemize}[leftmargin=*,noitemsep]
    \vspace{-0.25em}
    \item \textbf{\ic{<equal, *, *, *>}} captures \ic{equal} operator and applies its bypass logic at entry, i.e., fast comparison on two local Tensors with a boolean output, rather than go through Sharding Propagation for slow comparison on two DTensors with unnecessary inference overhead for output metadata.
    \item \textbf{\ic{<*, out-given, *, *>}} captures any operators with output tensor already given (placeholder to fill values), such as \ic{mm(a, b, out=c)} in Figure~\ref{fig:dtensor_dispatch_opt}, and directly uses \ic{c}'s metadata. % if \ic{c} exists.
    \item \textbf{\ic{<*, *, *, out-partial>}} captures any operators with output placement known to be \ic{Partial}, i.e., in half-way of Sharding Propagation, and bypasses the rest inference and uses produced local output's metadata plus \ic{Partial}.
    \item \textbf{\ic{<add, *, partial-replicate, *>}} captures \ic{add(a, b)} operator with \ic{a} in \ic{Partial} and \ic{b} in \ic{Replicate} and applies \textit{communication-free} add, i.e., convert \ic{b} into \ic{Partial} by zero-out \ic{b} on non-first devices, and then add legally with \ic{a}, which would otherwise requires an AllReduce~\cite{thakur2005mpi} to communicate \ic{a} into \ic{Replicate} before adding legally with \ic{b}.
    \vspace{-0.25em}
\end{itemize}
All rules follow \textit{first-match} principle and are ordered properly, and they can also be extended or customized by users. 

\para{\circled{2} Sharding Propagation Cache.}
We employ a caching mechanism that tracks the all operators and their input metadata.
This allows \vescale to reuse previous inferred results and skip the rest of Sharding Propagation.
Although native DTensor dispatch has caching, its hashing overhead is too prohibitive, due to its high complexity of $O(M \cdot E)$ where $M$ is number of metadata and $E$ is number of operands.
Therefore, we develop a light-weight hashing to minimize $M$ and $E$ while still avoid collision.
Furthermore, although cache misses do happen when training with real data of variable sequence lengths, our caching remains effective, as LLMs enjoy repetitive module structures.

\para{\circled{3} Efficient C++ Core.}
Despite the aforementioned optimizations, there remains 60+ $\mu$s overhead per dispatch, which stems from essential compute around the cache (unwrapping operands, parsing metadata, computing hash values, etc.).
To further reduce these Python-induced overheads, we implement critical components in C++ for faster Sharding Propagation.

\para{\circled{1}+\circled{2}+\circled{3}.}
With all three optimizations, \vescale cuts around 95\% of DTensor dispatch overhead and limits the overhead to approximately 30 $\mu$s for all operators (see Table~2 in the supplementary material for more details).

\subsection{Achieving Zero Dispatch Overhead}
\label{subsec:zero_dtensor_dispatch}
\para{Motivation.} 
Despite the success of \vescale's Eager mode in our production, we encountered scenarios where even minor dispatch overhead per operator is intolerable, particularly in models containing Mixture of Experts (MoE).
Because thousands of lightweight operators can be bounded by total CPU overhead of dispatching instead of the GPU kernel time. 

To achieve zero dispatch overhead, we introduce a new execution mode called \textit{Static Eager} mode.
The key insight is that the metadata of DTensors remain mostly static during runtime. 
For example, the weight DTensor remains \ic{Shard} placements throughout tensor parallel training.
The output activation DTensors remain the same placements of input DTensors for element-wise or inplace operators.
In practice, only few critical places really need dynamics in metadata of DTensor, such as the DTensor redistribution (recall Figure~\ref{fig:dtensor}) enforcing placement change, or the non-sharded weight gradient with a dynamic placement during backward.
Luckily, these places appear much less frequently compared to the total number of operators in a training iteration.
Therefore, we can treat DTensors' metadata as static and eliminate the entire DTensor dispatch to avoid the expensive metadata inference.
Only for those dynamic places, we handle them specially.

\para{Static Eager Mode.} 
Under this mode, \vescale discards the DTensor abstraction at runtime and executes all operators directly on local tensors, leaving only \textit{Local Tensor Execution} in Figure~\ref{fig:dtensor_dispatch_opt}.
For the aforementioned dynamics, \vescale still avoids DTensors by replacing inferred dynamic metadata with \textit{pre-described} static metadata.
This is achieved via \vescale's Plan API that can pre-describe the metadata, and then at initialization (before runtime) convert them into PyTorch hooks.
At runtime, those hooks are called mechanically to execute same logic as if DTensor still exist.
To be specific:
\begin{itemize}[leftmargin=*,noitemsep]
\vspace{-0.25em}
    \item \textbf{\ic{plan.redistribute(path, src, dst)}}: facilitates DTensor redistribution of a target tensor with a pre-described
 source and destination placement during forward. 
    For example, \ic{plan.redistribute(matmul.<in>, Shard(0), Replicate)} realizes the redistribution stage in Figure~\ref{fig:dtensor}. 
    Before runtime, the redistribution is lowered into Plan IR, abbreviated as \ic{matmul.forward\_pre:<in>:0:redist(S(0)->R)}.
    Then the IR is converted into a forward-pre hook registered at \ic{matmul}.
    During forward, the hook is called to execute AllGather communication, redistributing the first input tensor from sharded into replicated.
    If needed, the redistribution of gradients during backward can also be implemented by \ic{plan.redistribute(..., grad\_src, grad\_dst)}.
    
    \item \textbf{\ic{plan.annotate(path, placement)}}: annotates a target local tensor with a pre-described placement, as if it were a DTensor.
    For example, \ic{plan.annotate(weight.grad, Partial)} labels the gradient of \ic{weight} (e.g., a replicate weight in sequence parallel) to be \ic{Partial} and let it wait for a pending gradient reduction.
    It is also lowered into Plan IR and then converted into a backward-post hook which correctly calls gradient reduction.
    Furthermore, this API can also annotate activations for certain operators demanding DTensor placement.
    E.g., \ic{plan.annotate(dropout.<in>, Shard(0))} labels the input tensor's placement for a dropout such that it can trigger the distributed RNG logic (\S\ref{sec:correctness}).
\vspace{-0.25em}
\end{itemize}
Leveraging the Plan APIs above, Static Eager can reproduce the entire training with DTensors in Eager mode but achieve zero dispatch overhead for nearly all operators.

\para{Static Eager Plan Generation.}
While Static Eager improves the overall performance (4.62$\times$ in \S\ref{subsec:ablation}), planning \ic{redistribute} and \ic{annotate} demands some knowledge of parallelisms. 
To remove this demand, \vescale can automatically generate these Plans by leveraging warm-up iterations during regular training or a pre-run with Fake Tensors~\cite{torch2025faketensor}.

\subsection{Optimized DTensor Communication}
\label{subsec:optimize-comm}

\para{DTensor Bucketed Gradient Reduce.}
While DTensor provides an elegant abstraction for eager SPMD, it is fundamentally limited by this abstraction in that its communication (DTensor redistribution) must only occur at per-tensor granularity.
This limitation is most dire when meeting gradient reduce, a crucial communication in distributed training that requires reducing weight gradients of entire model across all devices (e.g., AllReduce~\cite{thakur2005mpi}). 
Because weight gradients are also DTensors, meaning gradient reduction (redistribution from \ic{Partial} to \ic{Replicate} via AllReduce) must also occur at per-gradient granularity, resulting in numerous small communication pieces.
As known to the network community, such fine-grained communication is detrimental to the bandwidth utilization. 
Hence, it is tempting to use the classic DDP~\cite{li2020pytorch} for bucketing multiple gradients together for coarse-grained reduction.
However, these works lack generality and can only handle data parallel training, ignoring other parallelisms that also require gradient reduction, such as sequence parallel~\cite{korthikanti2023reducing} and context parallel~\cite{liu2023ringatten, jacobs2023ulysses} which are essential for LLMs.

Therefore, we develop a \textit{general gradient reduce that can bucket together DTensors} for each parallel strategy (data/sequence/context/etc.), such that all gradient communications are bucketed to boost the bandwidth utilization.
This is achieved by: 
1) find DTensors with \ic{Partial} placements on the same dimension of the device mesh,
2) extract their local tensors to fill a bucket-sized communication buffer,
3) AllReduce,
4) reverse step 2 and change placements to \ic{Replicate},
and 5) repeat above for all dimensions of device mesh.

\begin{figure}[t]
    \centering
    \includegraphics[width=1\linewidth]{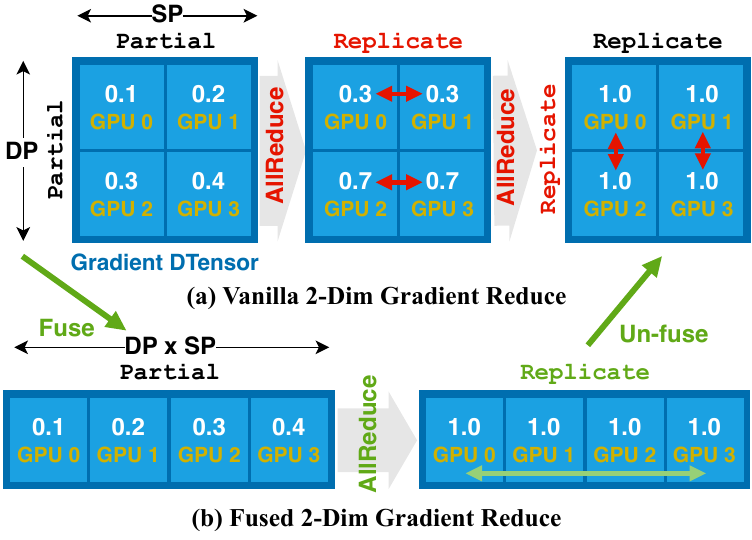}
    \vspace{-1em}
    \caption{\vescale's N-Dim fused gradient reduce. 
    Illustrated by a toy example of a single DTensor with \ic{Partial} placements on a 2-Dim device mesh of GPUs, where row and column GPUs are for sequence parallel (SP) and data parallel (DP), respectively. 
    The goal is to communicate each of the local tensors (e.g., $0.1\sim0.4$) and sum all via AllReduce.}
    \vspace{-1em}
    \label{fig:nd_fused_grad}
\end{figure}

\para{N-Dim Fused Gradient Reduce.}
Beyond bucketing multiple DTensors, we find an another opportunity to further increase the granularity of communication -- \textit{fusing the N-dimensional parallelisms into one}.
Specifically, recall each DTensor represents a global Tensor sharded on a N-Dim mesh of devices with each dimension used for one training parallelism.
When comes to gradient DTensor and reduction, the AllReduce must occur at per-dimension granularity, i.e., occur individually along each dimension of the mesh and repeat for all N dimensions.
Figure~\ref{fig:nd_fused_grad}(a) shows a 2-Dim example.
Therefore, we develop a \textit{N-Dim gradient reduce that fuses multiple parallel dimensions into one} for a single coarse-grained communication, further improving the bandwidth, as shown in Figure~\ref{fig:nd_fused_grad}(b).
This is achieved by: 
1) find dimensions of \ic{Partial}, 
2) fuse those dimensions into one by flattening the (sub) device mesh,
3) AllReduce on the flatten mesh,
and 4) change placements to \ic{Replicate} and mesh to the original one.

\input{table/eval_loc}

Analytically, we validate the benefit of N-Dim fused reduce by adopting the standard cost model of Ring AllReduce~\cite{thakur2005mpi}.
The vanilla N-Dim reduce takes the cost of:
$$T_{vanilla} \propto 2 S B \cdot \sum_{i=0}^{N-1}\frac{P_i - 1}{P_i} \approx 2 S B \cdot \textit{\bf N}$$
where $S$ is the size (a local tensor size or bucket size) in byte, 
$B$ is transfer time per byte\footnote{$B$ is roughly similar for intra-node interconnects and inter-node links in modern training clusters~\cite{microsoft2024h100cluster,amazon2025p5cluster,nvidia2025dgxh100}. 
E.g., each GPU in Azure H100 cluster can send message over intra-node NVLinks with 450 GBps per direction and also over inter-node InfiniBand links with 400 GBps (8x400 Gbps) per direction, as modern NCCL network stack can utilize all NICs for each GPU.
}, 
$N$ is number of dimensions demanding reduce, 
and $P_i$ is device counts along $i$-th dimension\footnote{The network latency term is omitted as it is negligible compared to bandwidth term $B$ in LLM training where bandwidth and throughput dominate.}.
By contrast, the fused N-Dim reduce costs:
$$T_{fused} \propto 2 S B \cdot \frac{ \prod_{i=0}^{N-1} P_i - 1}{\prod_{i=0}^{N-1} P_i} \approx 2 S B \cdot \textit{\bf 1} $$
Thus, for distributed training at scale (large $P$), the vanilla gradient reduce suffers from $N\times$ the communication cost while the fused gradient reduce enjoys a cost of only $1\times$.

%% file: table/eval_loc.tex
\begin{table*}[t]
\caption{Line of Code (LoC) change needed to adapt a single-device training script to 4-Dim parallel training.}
\label{tab:eval_loc}
\vspace{-1em}
\centering

\begin{adjustbox}{width=1\linewidth}

\begin{tabular}{|c|cccc|cccc|cccc|}
\hline
\multirow{2}{*}{System} & \multicolumn{4}{c|}{\llama-3}                                                                       & \multicolumn{4}{c|}{\mixtral}                                                                        & \multicolumn{4}{c|}{LI-\dit}                                                                        \\ \cline{2-13} 
                        & \multicolumn{1}{c|}{Model} & \multicolumn{1}{c|}{Plan} & \multicolumn{1}{c|}{Runtime} & Total  & \multicolumn{1}{c|}{Model} & \multicolumn{1}{c|}{Plan} & \multicolumn{1}{c|}{Runtime} & Total     & \multicolumn{1}{c|}{Model} & \multicolumn{1}{c|}{Plan} & \multicolumn{1}{c|}{Runtime} & Total \\ \hline
\megatron               & \multicolumn{1}{c|}{126}    & \multicolumn{1}{c|}{0}    & \multicolumn{1}{c|}{0}         & 126     & \multicolumn{1}{c|}{162}    & \multicolumn{1}{c|}{0}    & \multicolumn{1}{c|}{0}         & 162        & \multicolumn{1}{c|}{82}    & \multicolumn{1}{c|}{0}    & \multicolumn{1}{c|}{0}         & 82    \\ \hline
\deepspeed              & \multicolumn{1}{c|}{126}    & \multicolumn{1}{c|}{4}    & \multicolumn{1}{c|}{0}         & 130     & \multicolumn{1}{c|}{162}    & \multicolumn{1}{c|}{4}    & \multicolumn{1}{c|}{10}         & 176        & \multicolumn{1}{c|}{82}    & \multicolumn{1}{c|}{4}    & \multicolumn{1}{c|}{0}         & 86    \\ \hline
\titan                  & \multicolumn{1}{c|}{6}     & \multicolumn{1}{c|}{57}   & \multicolumn{1}{c|}{0}         & 63     & \multicolumn{1}{c|}{6}     & \multicolumn{1}{c|}{79}   & \multicolumn{1}{c|}{15}        & 100        & \multicolumn{1}{c|}{8}     & \multicolumn{1}{c|}{48}   & \multicolumn{1}{c|}{39}        & 95    \\ \hline
\textbf{\vescale}       & \multicolumn{1}{c|}{\bf 0}     & \multicolumn{1}{c|}{\bf 29}   & \multicolumn{1}{c|}{\bf 0}         & \bf 29 & \multicolumn{1}{c|}{\bf 0}     & \multicolumn{1}{c|}{\bf 38}   & \multicolumn{1}{c|}{\bf 0}         & \bf 38    & \multicolumn{1}{c|}{\bf 2}     & \multicolumn{1}{c|}{\bf 36}   & \multicolumn{1}{c|}{\bf 0}         & \bf 38 \\ \hline
\end{tabular}

\end{adjustbox}
\vspace{-1em}

\end{table*}

%% file: section/7-Implementation.tex
\section{Implementation}
\label{sec:implementation}

\vescale is implemented with 76K lines of code (LoC) in Python, 1K LoC in C++, and 300 LoC in CUDA. 

\para{Parallelized Model.} As discussed in \S\ref{sec:overview}, it is implemented with PyTorch-native hooks on \ic{nn.Module}\cite{pytorch2025modulehook} and \ic{torch.Tensor}\cite{pytorch2025tensorgradhook} to achieve automatic parallel in eager execution, without intrusively coupling model definition. 

\para{Distributed RNG.} Algorithm~\ref{alg:rng} is implemented in CUDA and seamlessly plugged into PyTorch library to support various random operators such that 
consistent distributed RNG is invoked transparently at runtime during DTensor dispatching.

\para{DTensor Extension.} We extend PyTorch DTensor for more functionalities to meet production demands, such as runtime creation of leaf DTensor, a new placement \ic{InterleavedShard} for non-contiguous shard, and deferred initialization without OOM. 
Details can be found in
\S~A
in supplementary material.

%% file: section/8-Experiments.tex
\section{Evaluation}
\label{sec:evaluation}

\subsection{Experimental Setup}
\label{subsec:setup}

\para{Environment.}
All experiments were conducted in a cluster of 32 servers with each server equipped with 8 H100-80~GB GPUs, 88 CPU cores, and 1.8 TB CPU memory. 
Within each server, GPUs are connected via NVSwitch with 450 GBps bandwidth for GPU-to-GPU communication. 
Servers are interconnected using InfiniBand with 8x400 Gbps links. 
Software stack uses PyTorch 2.4, NCCL v2.19, and CUDA v12.2.

\para{Models.}
We evaluated three types of large models with dense and sparse structures for language and vision tasks:
\begin{itemize}[leftmargin=*,noitemsep]
  \vspace{-0.25em}
  \item {\bf \llamathree}~\cite{dubey2024llama} of {\bf 8B}~\cite{llamathree} and {\bf 70B}~\cite{llamatwo70b}
  model size: a dense language model stacked by homogeneous Transformers.
  \item {\bf \mixtral}~\cite{jiang2024mixtral} of {\bf 3B} and 56B {\bf (8$\times$7B)}~\cite{mixtral8x7b} model size: a sparse language model with Mixture-of-Experts (MoE). 
  \item {\bf LI-\dit} of {\bf 10B}~\cite{ma2024exploring} model size: a diffusion model consisting of both Convolutions and Transformers.
  \vspace{-0.25em}
\end{itemize}
All models are trained with Adam optimizer on real datesets with a sequence length of 4K.
The default precision is BF16.

\para{Baselines.} 
We compare with three prominent training systems in the PyTorch ecosystem:
\begin{itemize}[leftmargin=*,noitemsep]
    \vspace{-0.25em}
    \item {\bf \megatron}~\cite{shoeybi2019megatron}: a widely adopted system dedicated for LLM training, known for its efficiency in tensor and sequence parallel but missing optimizer parallel ZeRO-3.
    \item {\bf \deepspeed}~\cite{mdeepspeed2022github}: an enhanced system on top of the previous one with extra ZeRO-3 of DeepSpeed~\cite{rajbhandari2020zero}.
    \item {\bf \titan}~\cite{torchtitan2024}: a popular PyTorch-native system for LLM training, known for its Eager SPMD with DTensors for tensor, sequence, data parallel and ZeRO-3.  
    \vspace{-0.25em}
\end{itemize}
For fair comparison, all baselines and \vescale share the same 4-Dim parallel: tensor, sequence, data, and ZeRO-3 (except \megatron's ZeRO-1).
Within tensor and sequence parallel, the specific shardings are aligned to \megatron.

\subsection{Development Effort in Lines of Code (LoC)} 
\label{subsec:exp-loc}
We measure the development effort needed to adapt a single-device training script for 4-Dim parallel training by how many LoCs are changed.
The single-device scripts are sourced from HuggingFace~\cite{wolf2019huggingface} for \llama and \mixtral, and the official repository for LI-\dit~\cite{peebles2023scalable}. Table~\ref{tab:eval_loc} compares the results. 

First, we observe that \vescale achieves \textit{almost zero change on all model code}
\footnote{We truthfully report the 2 LoC change in LI-\dit, which is to fix the model's inconsistent tensor size between initialization and runtime. This inconsistency is hard coded and is to be reported as a potential model bug.} 
with fully decoupled distributed logic, thus enjoying simple development of models as if on a single device, without worrying about intricate distributed systems.
By contrast, baselines like \megatron require both heavy and intrusive changes on original model definition for different parallelisms (e.g., \ic{ColumnParallelLinear} for tensor parallel \ic{Linear}) with up to 162 LoCs, which incurs not only extra effort from system experts but also repeated effort across different models and different parallelisms.

Second, \vescale only needs \textit{little effort to describe parallelisms} in the Plan by leveraging the Plan APIs (\S\ref{sec:api}).
Such effort from users is minimal, e.g., 50\% less LoC than \titan that also uses descriptive plans.
The advantages of \vescale Plan API are threefold:
1) its generality that does not bind a specific parallelism to a certain operator, while \titan plan binds \ic{RowwiseParallel} to \ic{Linear}, thus failing to support new operators or new parallels unless users develop their own (e.g., MoE operators or hybrid parallel);
2) its automation that does not demand users' knowledge of source sharding placement nor ask users' decision if needing Tensor-DTensor conversion at each operator, while \titan plan relies on both due to its inherent optimization using mixed Tensor and DTensor;
and 3) its expressiveness that matches tensors with regular expression patterns, while \titan requires manually traversing a model to check each tensor before describing the plan, even for repeated model structures like LLMs.

Third, \vescale also requires \textit{zero change on its runtime} to support diverse model architecture like MoE and Diffusion. 
This is not possible without extending the runtime in both \deepspeed and \titan, since the former misses ZeRO-3 support for expert weights and the latter currently lacks tensor parallelism for expert modules.

Consequently, \vescale enjoys the minimal development effort among all baselines, saving up to 78.4\% total LoC.

\begin{figure}[!t]
\centering    
    \begin{subfigure}[b]{0.49\linewidth}
         \centering
         \includegraphics[width=\linewidth]{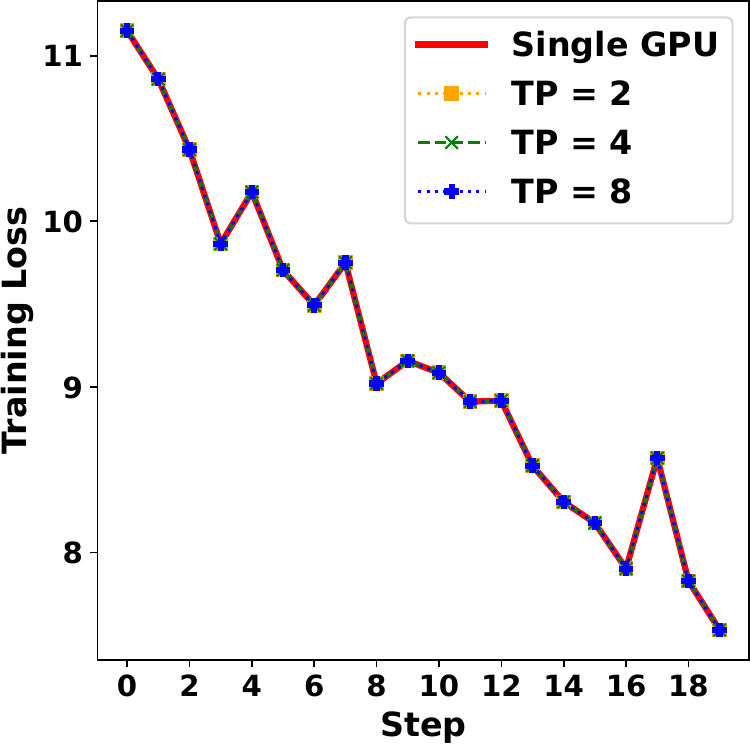}
         \vspace{-1.5em}
         \caption{Random initialize only}
    \end{subfigure}
    \hfill
    \begin{subfigure}[b]{0.49\linewidth}
         \centering
         \includegraphics[width=\linewidth]{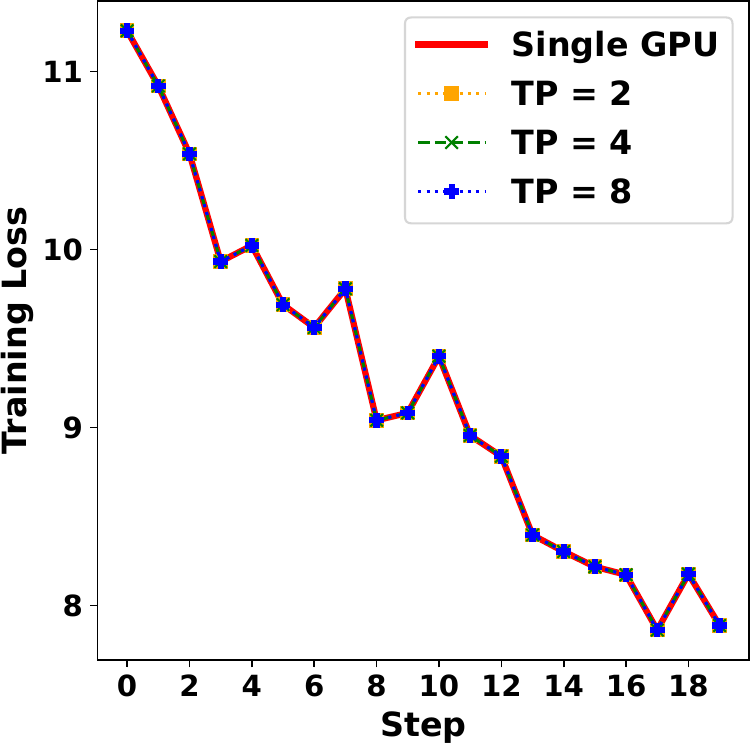}
         \vspace{-1.5em}
         \caption{Random dropout only}
    \end{subfigure}
    \vspace{-1em}
    \caption{\vescale's single-device semantics in terms of training loss of \llamathree-8B with tensor parallel and FP32.}
    \label{fig:eval_vescale_rng}
    \vspace{-1em}
\end{figure}

\subsection{Single-Device Semantic}
We evaluate the semantics of SPMD training in two cases:
\begin{itemize}[leftmargin=*,noitemsep]
  \vspace{-0.25em}
  \item {\bf Random initialization only}: only allow randomness in distributed weight initialization and remove other randomness like dropout operators during training.
  \item {\bf Random dropout only}: only allow randomness in distributed dropout operators and remove other randomness like weight initialization.
  \vspace{-0.25em}
\end{itemize}
For fair comparison, all baselines and \vescale share the same initial random state, data sampling, model definition, and training hyper-parameters.

\para{Training Convergence.}
Figure~\ref{fig:eval_vescale_rng} compares training loss of \vescale between various tensor parallelism sizes and the single device.
We observe that \vescale achieves exact alignment with single-device results across different device numbers in both cases of random initialization and random dropout,
which is attributed to the proposed distributed RNG with virtualized threads (\S\ref{sec:correctness}).
In contrast, all baselines (\titan, \megatron, and \deepspeed) exhibits significant deviations from the single-device result in both cases, up to \textit{four orders of magnitudes} (recall Figure~\ref{fig:related_work_incorrectness} and see \S~B in the supplementary material for more details). 
Especially, such deviations vary substantially across different device numbers, leading to \textit{inconsistent semantics of multi-device execution}, let alone single-device semantic.
Such inconsistent semantics root from either seed-based RNG or offset-based RNG with mismatched seeding or offseting in non-first devices (see Figure~\ref{fig:rng_compare}(d) and (e)). 

\input{table/eval_correct}

\para{Numerical Difference.}
All baselines show significant divergence, with maximal differences up to \textit{0.9299}, \textit{1.1038}, and \textit{1.5870}, respectively. 
In contrast, \vescale incurs negligible maximal, under \textit{6e-5}, which stems primarily from the inevitable differences in collective communications like AllReduce, rather than randomness.
As known to the community, AllReduce reorders addition operations of a single device and floating-point addition is not associative~\cite{stackoverflow2012fpaddassociate, nvidia2023allreducesum}, thus the inherent numerical difference in parallel training.
Such minor differences can be amplified drastically under inconsistent RNG, making debugging and diagnosing extremely difficult.

\para{Micro-benchmark.}
Furthermore, we evaluate per-operator randomness by comparing DTensor to single-device tensor, both generated from the same random operator. 
We observe that \vescale's DTensor is bit-wise equal to single-device tensor across all common operators (\ic{normal}, \ic{uniform}, \ic{Dropout}, etc.) under every combination of tensor dimensions (1D$\sim$5D), sizes (KB$\sim$GB), and shardings (\ic{Shard(dim=0\~4)}).
Details can be found in \S C in the supplementary material.

\begin{figure*}[!t]
    \centering
    \includegraphics[width=1\linewidth]{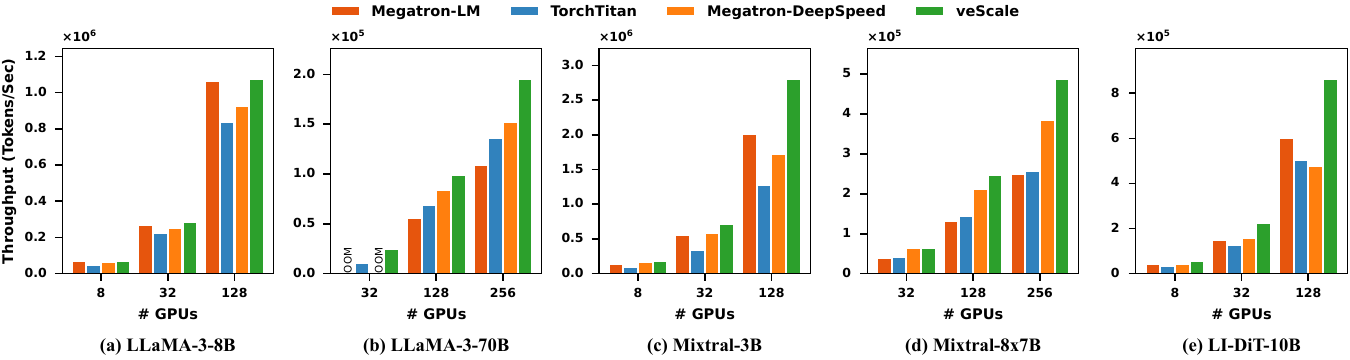}
    \vspace{-2em}
    \caption{End-to-end training throughput with 4-Dim parallel (tensor, sequence, data, and ZeRO) and up to 256 GPUs.}
    \vspace{-1em}
    \label{fig:throughput}
\end{figure*}

\subsection{End-to-End Performance}
Next, we evaluate end-to-end training performance. 
For a fair comparison, all baselines and \vescale share the 4-Dim parallel (\S\ref{subsec:setup}) and the same global batch size, with tuned per-parallel size and per-device batch size to maximize performance individually for each system. 
Also, optional plugin of CUDA kernels (esp., \megatron's GroupedGEMM~\cite{nvidia2024groupedgemm} for replacing PyTorch-native kernels in MoE layers) is disabled to align all systems, as plugin CUDA kernels are not the focus of this evaluation and in fact are complementary to all systems.
Figure~\ref{fig:throughput} summarizes the comparison.

First, we observe that \vescale demonstrates impressive performance, outperforming all baselines in all cases across different model types, models sizes, and numbers of GPUs.
For example of \llamathree-70B, \vescale is up to \textit{1.8$\times$, 1.4$\times$, and 1.3$\times$ faster} than \megatron, \titan, and \deepspeed, respectively.
Comparing \vescale with \megatron, the former shows better performance primarily due to its support for ZeRO-3~\cite{rajbhandari2020zero} that fully shards parameters, gradients, and optimizer states within data parallel. 
In contrast, \megatron is limited to ZeRO-1, where only optimizer states can be sharded. 
This results in increased memory usage for \megatron, necessitating larger tensor parallelism sizes that ultimately hinder efficiency. 
For \llamathree-70B, \megatron requires 8-way tensor parallelism to support a batch size of only 1 for each GPU. 
Conversely, all other systems (\titan, \deepspeed, and \vescale) support ZeRO-3 and thus only require 4-way tensor parallelism to accommodate a larger batch size of 2.
The advantage of \vescale over \titan is largely attributed to the reduced DTensor CPU overhead as well as optimized communication~(\S\ref{sec:perf_optimization}). 
While \titan attempts to mitigate CPU overhead by converting DTensor back to local Tensors for certain operators like FlashAttention, it still experiencs significant overhead in other layers, such as Linear, Embedding, and LayerNorm. 
This inefficiency is compounded by the frequent conversions between DTensors and local Tensors, further impacting its overall performance. 
Comparing with \deepspeed, \vescale is still more efficient, mostly due to its highly optimized communication that not only buckets gradient reduce beyond data parallel but also fuses them across all related parallelisms~(\S\ref{subsec:optimize-comm}).

Secondly, we observe that \vescale enjoys a larger winning margin for training sparse models like \mixtral, with up to \textit{2.2$\times$ speedup over all baselines} for \mixtral-3B and up to 1.9$\times$ speedup for \mixtral-8x7B.
This notable improvement stems from the increased DTensor overhead inherent to the sparse architecture of MoE compared to dense architectures, which widens the gap between \titan and \vescale.
Meanwhile, the ZeRO-3 advantage over \megatron still holds, so does the improvement of \vescale's optimized communication over \deepspeed.

Lastly, for vision tasks, \vescale also provides up to \textit{1.8$\times$ speedup over all baselines} for LI-\dit-10B.

\begin{figure}[!t]
    \centering
    \includegraphics[width=1\linewidth]{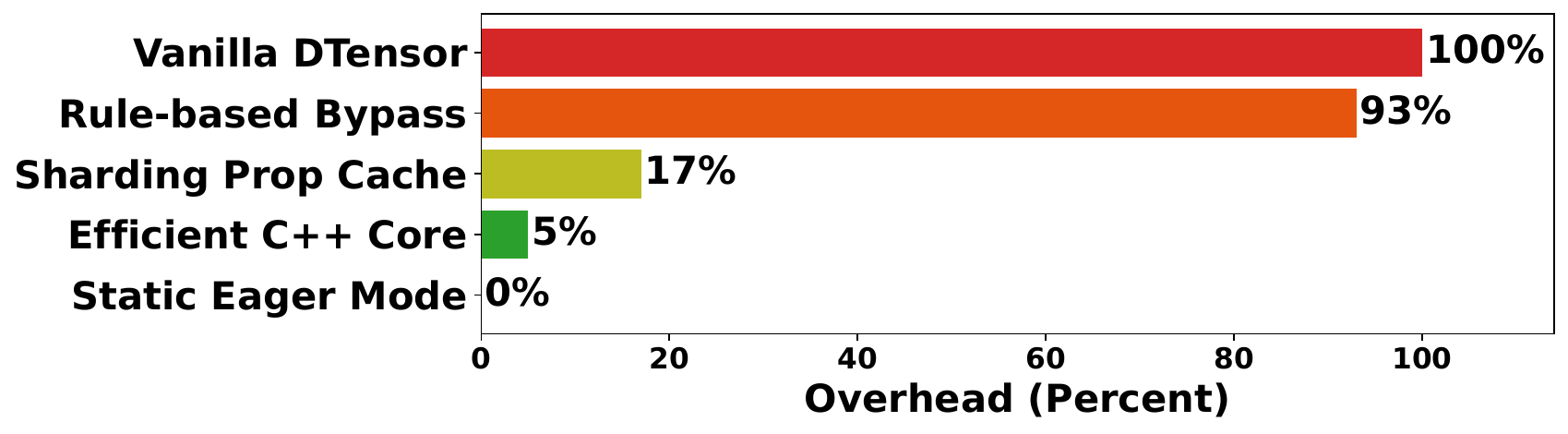}
    \vspace{-2em}
    \caption{Microbenchmark of DTensor overhead mitigation on training \mixtral-3B with tensor parallel on 8 GPUs.}
    \label{fig:ablation_dtensor}
    \vspace{-1em}
\end{figure}

\subsection{Ablation Study}
\label{subsec:ablation}

To understand the speedup, we conducted ablation study on \vescale's optimizations against the vanilla PyTorch DTensor without any optimizations even from \titan. 

\para{DTensor Overhead Mitigation.} 
Figure~\ref{fig:ablation_dtensor} illustrates the benefits of various optimization techniques~(\S\ref{subsec:dtensor_dispatch_perf_optimization}, \S\ref{subsec:zero_dtensor_dispatch}) in reducing DTensor CPU overhead.
By incorporating rule-based bypass, \vescale achieves a 7\% reduction in overhead. 
Further optimization via sharding propagation cache contributes to an additional reduction of 76\% overhead. 
Even further, with the C++ core, the DTensor overhead is minimized to only 5\%. 
Ultimately, the overhead can be totally removed under Static Eager mode.
More details of per-operator overhead can be found in Table~2 in the supplementary material.
These reductions highlight the effectiveness of our approach in reducing CPU overhead, which in turn unblocks the full power of GPUs and contributes to the overall speedup.
We believe the CPU overhead will be even more severe in near future, considering the trend of ever-faster GPUs over CPUs~\cite{cornel2025cpuvsgpu}.

\begin{figure}[!t]
    \centering
    \includegraphics[width=\linewidth]{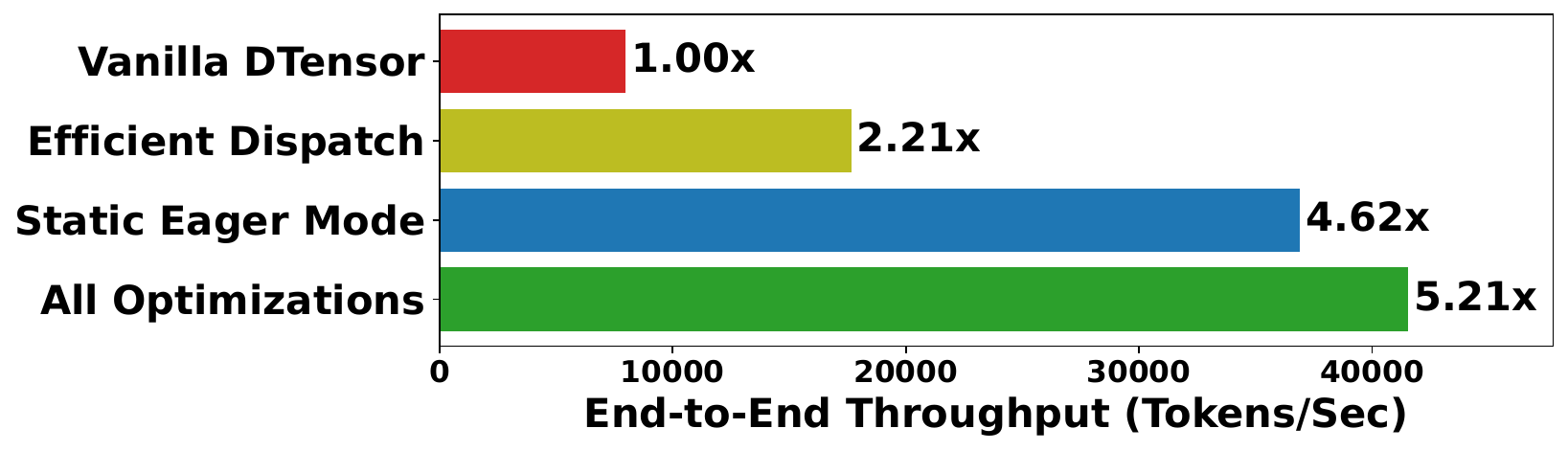}
    \vspace{-2em}
    \caption{Optimizations on training \mixtral-3B with 16-way sequence parallel and 4-way ZeRO-3 on 64 GPUs.}
    \label{fig:ablation_perf}
    \vspace{-1em}
\end{figure}

\para{All Optimization Breakdown.} 
To analyze the benefits of all optimizations including DTensor and communication (the entire \S\ref{sec:perf_optimization}), Figure~\ref{fig:ablation_perf} breaks down end-to-end training throughput into each of the optimizations.
Efficient dispatch along boots the throughput by $2.21\times$ over the vanilla DTensor.
Further plugging in Static Eager gives another $2.09\times$ speedup.
Lastly, turning on optimized communication and with all optimizations combined, 
\vescale push the speedup to $5.21\times$. 
Such a sharp improvement can be credited to not only the DTensor overhead mitigation but also the gradient reduce with tensor bucketing and parallel fusion (\S\ref{subsec:optimize-comm}).

%% file: table/eval_correct.tex
\begin{table}[!t]
\caption{Maximal numerical difference in training loss between distributed and single-device training for \llamathree-8B with FP32.}
\label{tab:eval_correct}
\vspace{-1em}
\centering
\footnotesize
\setlength{\tabcolsep}{0.4em}
\begin{tabular}{|c|c|c|c|c|c|}
\hline
\multirow{2}{*}{Case}  & \multirow{2}{*}{TP} & Megatron & Torch & Megatron & \multirow{2}{*}{\textbf{\vescale}} \\ 
  &  & LM & Titan & DeepSpeed  &   \\ \hline
\multirow{3}{*}{
\begin{tabular}[c]{@{}c@{}}
Random\\
Initialize\\
Only
\end{tabular}} & 2    &      0.\bad{4}999  &  0.\bad{2}539 & \bad{1}.5870 & {0.0000\good{6}2}  \\ \cline{2-6} 
               & 4    &      0.\bad{9}299  &  0.\bad{3}315 & \bad{1}.1715 & {0.0000\good{3}7}  \\ \cline{2-6} 
               & 8    &      0.\bad{3}380  &  0.\bad{3}010 & \bad{1}.0783 & {0.0000\good{2}1}  \\ \hline
\multirow{3}{*}{
\begin{tabular}[c]{@{}c@{}}
Random \\
Dropout\\
Only
\end{tabular}} & 2    &     0.\bad{4}924   & \bad{1}.1038  & 0.\bad{5}639 & {0.0000\good{1}4} \\ \cline{2-6} 
               & 4    &     0.\bad{3}845   & 0.\bad{8}049  & 0.\bad{4}488 & {0.00000\good{7}}   \\ \cline{2-6} 
               & 8    &     0.\bad{2}551   & 0.\bad{6}023  & 0.\bad{4}510 & {0.0000\good{1}3} \\ \hline
\end{tabular}
\vspace{-1em}
\end{table}

%% file: section/9-Related_Work.tex
\section{Related Work}
\label{sec:related-work}

\para{Distributed Training.}
The common parallelisms of SPMD training are: tensor (TP), data (DP), sequence (SP), optimizer parallel (ZeRO).
TP~\cite{shoeybi2019megatron} shards weights across devices and compute with input tensor (replicated or sharded) locally on each device, incurring AllGather/ReduceScatter communication on input/output.
DP~\cite{li2020pytorch} replicates weights and compute sharded input tensor, followed by AllReduce on weight gradients.
SP~\cite{korthikanti2023reducing} is similar to DP but shards input along the tensor dimension of sequence (rather than data) and also AllReduces weight gradient.
ZeRO~\cite{rajbhandari2020zero} shards optimizer states, and AllGathers weights from sharded to replicate for compute, followed by ReduceScatter on weight gradient.
All parallelisms can be combined and put onto its dimension of a device mesh.

\para{Compiler-based Distributed Training.} 
Compiler-based systems such as Alpa~\cite{zheng2022alpa}, GSPMD~\cite{xu2021gspmd}, and JAX~\cite{jax2018github} enable SPMD training with automatic parallelization on a single-device model.
The most recent \textbf{\ic{torch.compile}}~\cite{ansel2024torchcompile} also belongs to this family.
However, they rely on the success of tracing the model graph, which is hardly guaranteed in the internal production.
This is caused by those non-traceable or non-compilable components that are commonly seen in model graphs, such as 
Python objects and functions,
3rd-party libraries like NumPy,
in-house CUDA/C operators,
custom communications,
callback hooks,
complex control flows and data dependencies~\cite{torch2025untracablegraphbreak,torch2024compileautograd,jax2024traceerror,tf2024knowissue}.
In case of trace failure, \ic{torch.compile} offers the \textit{graph break} mechanism to fall back to eager execution~\cite{torch2025untracablegraphbreak}, but its performance suffers from extra overheads and restrictions~\cite{torch2025graphbreaknospeedup,torch2024compileslowerthaneager}.
Furthermore, compiler often aggressively optimizes performance with complex transformations on model graphs and communications, and can thus hinder debugging and understanding~\cite{tf2024debug,jax2024debug,torch2025compiledebug}.

\para{Eager-based Distributed Training.} 
In contrast, eager systems offer not only flexibility without relying on model graphs, but also  interactivity with line-by-line execution and debug.
Hence, they still dominate the community~\cite{assembly2023torchpercent}.
The eager distributed training systems (e.g., \megatron~\cite{shoeybi2019megatron,mdeepspeed2022github}), however, still have much to learn from the compilers, such as the SPMD programming without coupling parallelization with model definitions. 
This inspired a newer system \slapo~\cite{chen2024slapo} that improves the usability of \megatron TP, by building a set of APIs on top of it.
For instance, \ic{shard(param)} and \ic{annotate(param)} are to shard a parameter into TP with annotated attributes.
Nonetheless, only TP-specific parameter is considered, missing not only sharding of activations and gradients but also more complex parallelisms (SP, ZeRO).
The recent \titan offers the state-of-the-art eager SPMD with DTensor and descriptive plan, but its plan still lacks generality for arbitrary operators and parallelisms, let alone automation and pattern matching. 
The inconsistency and overhead of DTensor are also critical blockers for the internal production.

\para{Distributed RNG.} RNG not only affects convergence but also complicates debug, test, and reproduce.
However, ensuring consistent RNG across devices is challenging due to inherent isolation of RNG stream on each device. 
Such inconsistency is even direr when mixing different parallelisms (TP, DP, etc.) with each adding its own randomness, and when resharding to different scales (e.g., 2 to 8 GPUs).
Most systems (\megatron, \rawdeepspeed, \slapo, \colossal, \jax~\cite{shoeybi2019megatron, rasley2020deepspeed, chen2024slapo, li2023colossal, jax2018github}) omits (to some extent) 
the consistency by using seed-based RNG to generate \textit{statistically uncorrelated} random sequence~\cite{nvidia2024pseudorandom,jax2024rng} on each device, ending up multi-device semantics and binding each training outcome with its own parallel and scale.
\titan~\cite{torchtitan2024} attempted improvement by offseting the RNG state to sample the global random sequence with a local stride, but still deemed it as \textit{impossible} to match the single-device RNG~\cite{dtensor2024rngissue}.

%% file: section/10-Conclusion.tex
\section{Conclusion and Future Work}

\vescale is an distributed training system that achieves single-device semantic and high performance within the emerging eager-SPMD paradigm of decoupled parallelization from model definition. 
\vescale studies the fundamental primitive DTensor, and addresses not only its inconsistency through a novel distributed RNG algorithm but also its inefficiency by reducing compute and communication overhead. 
Evaluations show that \vescale achieves $\sim2.2\times$ speedup compared to the state-of-the-art systems, while preserving consistent single-device semantics and saving $78.4\%$ development effort.

We are looking forward towards collaborating with the TorchTitan Team and upstreaming \vescale features to the official PyTorch repository, making the PyTorch Distributed stronger and the open source community better.

%% file: section/11-Acknowledgements.tex
\section{Acknowledgements}
We gratefully thank our alumni, Li Chen, Zimeng Huang, and Jiannan Wang, for their contribution to the distributed RNG, and our alumnus, Chengji Yao, for his contribution to the efficient DTensor dispatch.
Thanks to everyone in VeOmni (\url{https://github.com/ByteDance-Seed/VeOmni}) team for our tight collaboration: Qianli Ma, Yaowei Zheng, Zhelun Shi, Zhongkai Zhao, Bin Jia, Ziyue Huang, and Zhi Zhang.
Importantly, thanks to everyone in TorchTitan team for our great discussion and collaboration in the open source community: Wanchao Liang, Tianyu Liu, Chien-Chin Huang, Ke Wen, Xilun Wu, Wei Feng, Howard Huang, Will Constable, Andrew Gu, Iris Zhang, Less Wright, Junjie Wang, Yifu Wang, Sanket Purandare, and Gokul Nadathur.

%% file: section/appendix.tex
%%%%%%%%%%%%%%%%%%%%%%%%%%%%%%%%%%%%%%%%%%%%%%%%%%%
\section{\vescale's DTensor Functionality Extension}
\label{appendix:dtensor_extension}

\para{Runtime Creation of Leaf DTensors.} 
It is a common practice to create tensors (e.g., calling~\ic{torch.zeros}) directly in ~\ic{Module.forward()} rather than in~\ic{Module.__init__}, which miss-es the DTensor semantic (e.g., placement) on those tensors.
This leads to the execution of operators on a mixture of non-converted Tensors and DTensors, thus giving unexpected computation results.
\vescale develops a new API called \ic{plan.factory(path, placement)} and enables conversion of even runtime-created Tensor into DTensor with desired \ic{placement}, thus resolving this issue.

\begin{figure}[h]
    \centering
    \includegraphics[width=1.0\linewidth]{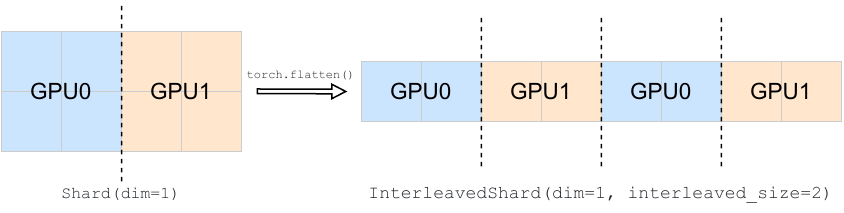}
    \caption{An example of \ic{InterleavedShard} usage.}
    \label{fig:placement_is}
\end{figure}

\para{New Placement -- \ic{InterleavedShard}.} 
The native DTensor placements are insufficient to cover all PyTorch operations.
For instance, operations like \ic{torch.flatten} on a sharded input tensor produce an output tensor that cannot be described by the existing placement types, as each device holds non-contiguous fragments of the global tensor.
Another common scenario occurs in the training of attention layers~\cite{vaswani2017attention} with tensor parallelism, where QKV parameters are fused into a single large parameter for matrix multiplication, resulting in non-contiguous fragments when sharding across attention heads.
To support these cases, \vescale introduces \ic{InterleavedShard} placement, which allows for sharding a dimension into smaller chunks arranged in an interleaved pattern. 
Figure~\ref{fig:placement_is} shows an example.

\para{Deferred Initialization} 
To support initializing giant models (e.g., with trillions of parameters), \vescale offers a \textbf{\ic{deferred\_init}} API for DTensor parameter initialization. 
Rather than immediately allocating memory for each DTensor, it records call stacks (e.g., uniform distribution) to each DTensor and delays memory allocation until \ic{parallelize}, applying it only to the necessary local tensor on each device. 
This approach accelerates initialization and prevents Out-Of-Memory (OOM) issues on both CPU and GPU.

\newpage

\section{Baselines' Semantic in Training Convergence}
\label{appendix:baseline_converge}
\begin{figure}[h]
\centering    
    \begin{subfigure}[b]{0.45\linewidth}
         \centering
         \includegraphics[width=\linewidth]{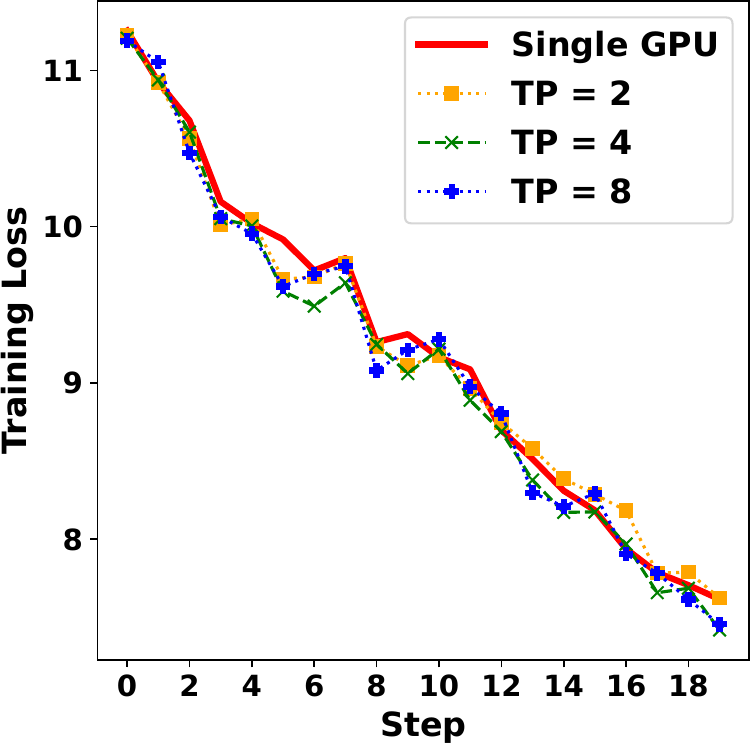}
         \caption{Random init. only}
    \end{subfigure}
    \hspace{0.5cm}
    \begin{subfigure}[b]{0.45\linewidth}
         \centering
         \includegraphics[width=\linewidth]{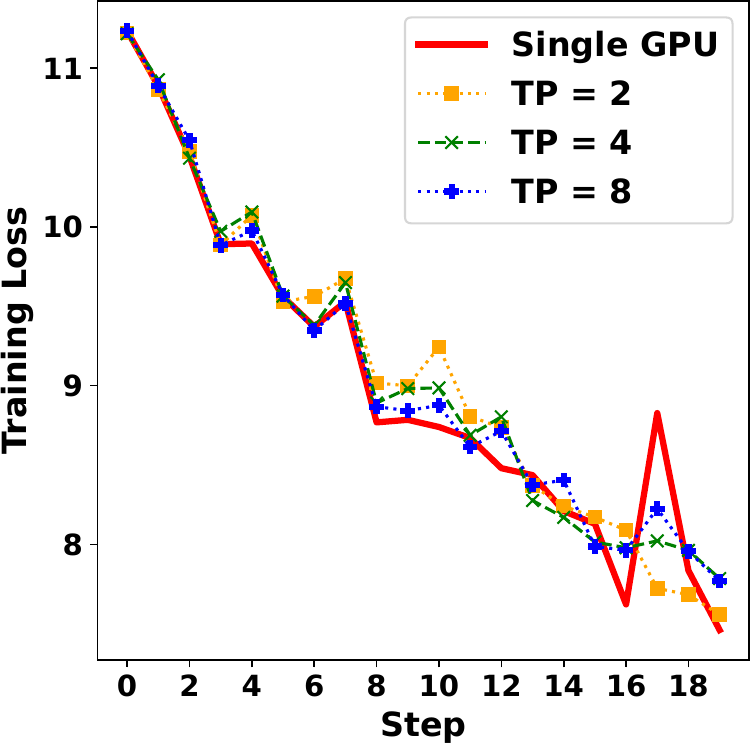}
         \caption{Random dropout only}
    \end{subfigure}
    \vspace{-2em}
    \caption{\titan's semantics in terms of training loss of \llamathree-8B with FP32.}
    \label{fig:eval_titan_rng}
    \vspace{-1em}
\end{figure}

\begin{figure}[h]
\centering    
    \begin{subfigure}[b]{0.45\linewidth}
         \centering
         \includegraphics[width=\linewidth]{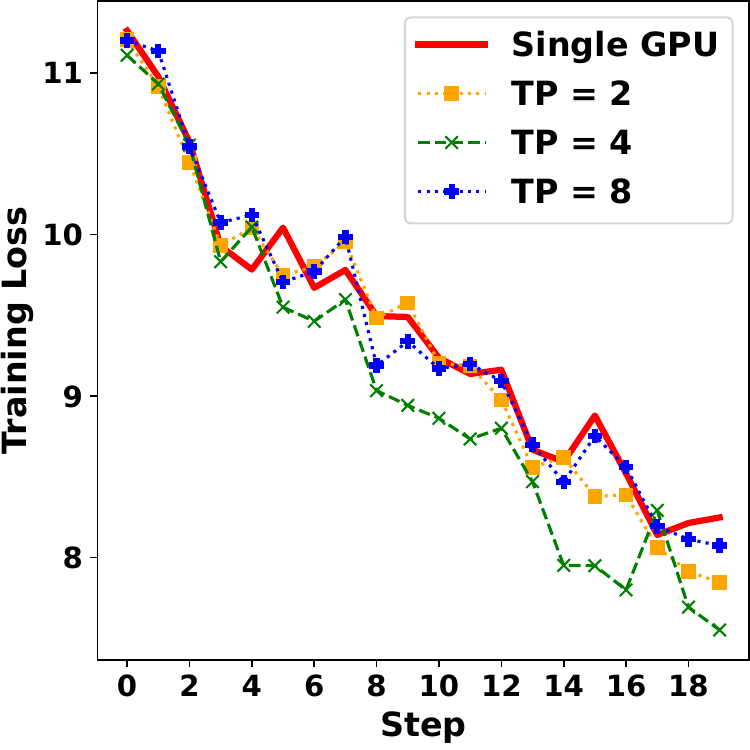}
         \caption{Random init. only}
    \end{subfigure}
    \hspace{0.5cm}
    \begin{subfigure}[b]{0.45\linewidth}
         \centering
         \includegraphics[width=\linewidth]{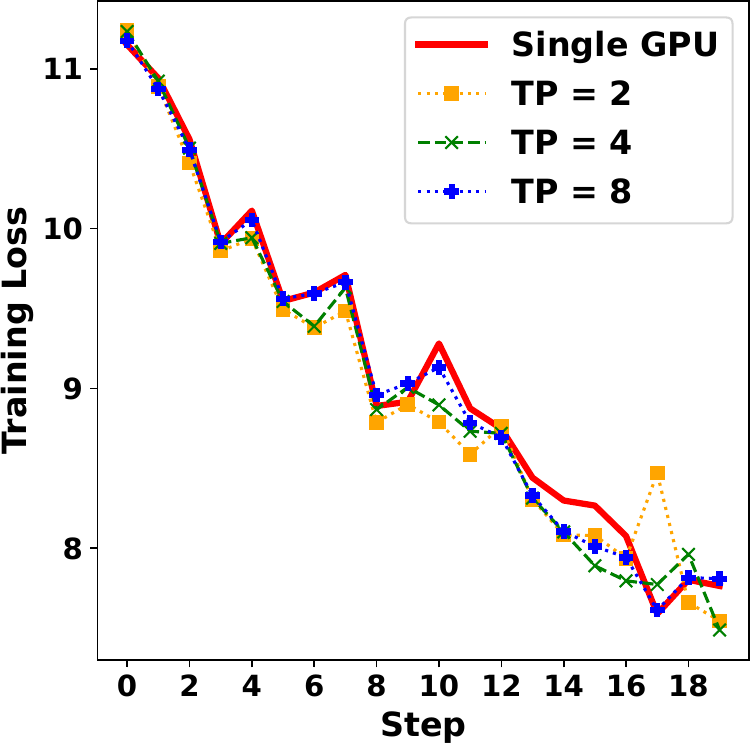}
         \caption{Random dropout only}
    \end{subfigure}
    \vspace{-2em}
    \caption{\megatron's semantics in terms of training loss of \llamathree-8B with FP32.}
    \label{fig:eval_megatron_rng}
    \vspace{-1em}
\end{figure}

\begin{figure}[h!]
\centering    
    \begin{subfigure}[b]{0.45\linewidth}
         \centering
         \includegraphics[width=\linewidth]{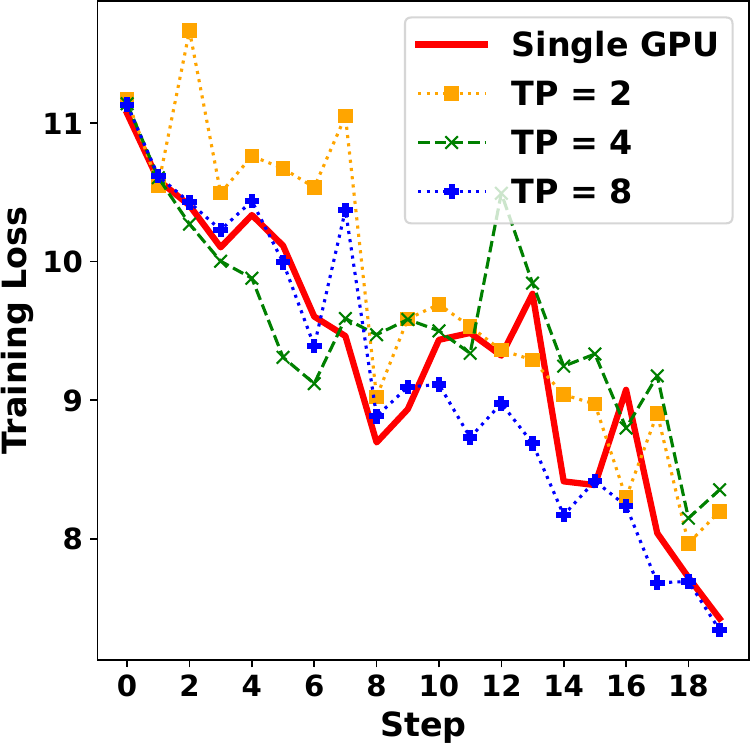}
         \caption{Random init. only}
    \end{subfigure}
    \hspace{0.5cm}
    \begin{subfigure}[b]{0.45\linewidth}
         \centering
         \includegraphics[width=\linewidth]{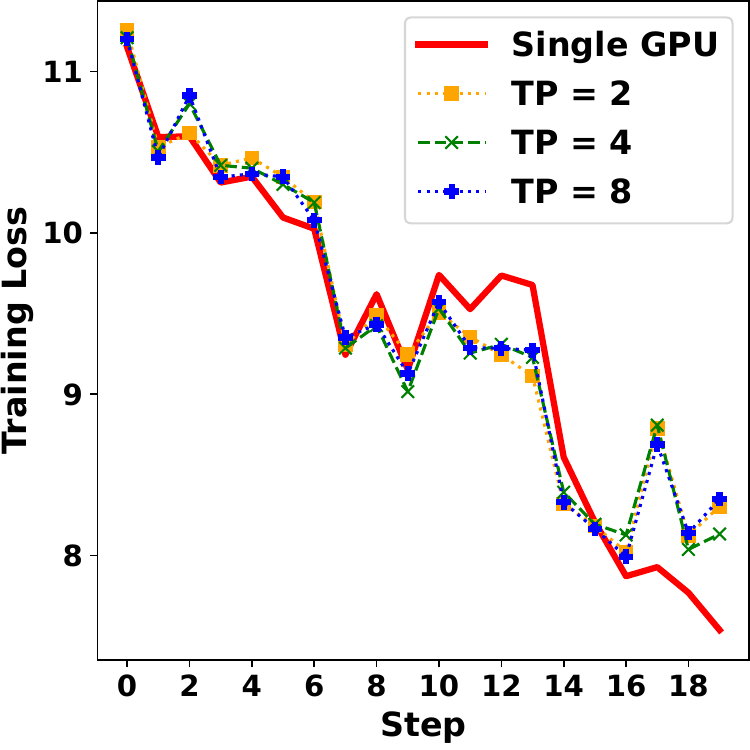}
         \caption{Random dropout only}
    \end{subfigure}
    \vspace{-2em}
    \caption{\deepspeed's semantics in terms of training loss of \llamathree-8B with FP32.}
    \label{fig:eval_deepspeed_rng}
    \vspace{-1em}
\end{figure}

Figure~\ref{fig:eval_titan_rng}~\ref{fig:eval_megatron_rng}~\ref{fig:eval_deepspeed_rng} show that neither \titan, \megatron, nor \deepspeed achieves consistent semantic for random weight initialization or random dropout in distributed training.

\newpage

\section{\vescale's Per-Operator Distributed RNG}
\label{appendix:rng}
Table~\ref{tab:rng_ops} evaluates the per-operator RNG where \vescale DTensor is bit-wise equal to single-device tensor.
E.g., bit-wise match is achieved for the \ic{nn.Dropout} generating ``1D'' to ``5D'' tensor with memory size from ``262KB'' to ``16GB'' under sharding on tensor dim ranging from first dim ``\#0'' to last dim ``\#4'' as well as doubling sharding on tensor dims ranging from ``\#0 and \#0'' to ``\#4 and \#4''.

\input{table/rng_ops}

\newpage

\section{\vescale's Efficient DTensor Dispatch in Each Operator}
\label{appendix:per_op_dispatch}
Table~\ref{tab:per_op_dispatch} breakdowns the DTensor dispatch overhead for each of the representative operators. 
The overhead is averaged across all occurrence for each operator.
The percentage is normalized per operator.

\input{table/per_op_dispatch}

%% file: table/rng_ops.tex
\begin{table}[h]

\caption{Per-operator randomness where \vescale DTensor is bit-wise equal to single-device tensor.
``Tensor Dims" refers to the number of dimensions, ``Tensor Size" shows the occupied memory size, and ``Sharded Dim" refers to the tensors dimensions to shard.
}
\centering
\label{tab:rng_ops}
\footnotesize
\begin{tabular}{|l|l|l|l|}
\hline
\begin{tabular}[c]{@{}l@{}}Random \\ Operator\end{tabular} &
  \begin{tabular}[c]{@{}l@{}}Tensor \\ Dims\end{tabular} &
  \begin{tabular}[c]{@{}l@{}}Tensor \\ Size\end{tabular} &
  \begin{tabular}[c]{@{}l@{}}Sharded \\ Dim\end{tabular} \\ \hline
\texttt{nn.init.normal\_} &
  \multirow{8}{*}{\begin{tabular}[c]{@{}l@{}}1D\\ $\sim$\\ 5D\end{tabular}} &
  \multirow{8}{*}{\begin{tabular}[c]{@{}l@{}}262 KB,\\ 64 MB,\\ 1 GB,\\ 16 GB\end{tabular}} &
  \multirow{8}{*}{\begin{tabular}[c]{@{}l@{}}\#0\\ $\sim$\\ \#4,\\ \#0 and \#0,\\ \#0 and \#1,\\ \#0 and \#2,\\ $\sim$\\ \#4 and \#3,\\ \#4 and \#4\end{tabular}} \\
\texttt{nn.init.uniform\_}                                                     &  &  &  \\
\texttt{\begin{tabular}[c]{@{}l@{}}nn.init.kaiming\_\\ uniform\_\end{tabular}} &  &  &  \\
\texttt{rand\_like}                                                            &  &  &  \\
\texttt{randn\_like}                                                           &  &  &  \\
\texttt{randint\_like}                                                         &  &  &  \\
\texttt{nn.Dropout}                                                            &  &  &  \\
\texttt{Tensor.uniform\_}                                                      &  &  &  \\ \hline
\end{tabular}

\end{table}

%% file: table/per_op_dispatch.tex
\begin{table}[h]
\caption{Per-operator DTensor overhead mitigation 
on training Mixtral-3B with tensor parallel on 8 GPUs}
\label{tab:per_op_dispatch}
\centering
\footnotesize
\begin{tabular}{|c|c|c|c|c|}
\hline
 &
  \begin{tabular}[c]{@{}c@{}}Matmul \\ \ic{mm}\end{tabular} &
  \begin{tabular}[c]{@{}c@{}}Transpose \\ \ic{t}\end{tabular} &
  \begin{tabular}[c]{@{}c@{}}View \\ \ic{view}\end{tabular} &
  \begin{tabular}[c]{@{}c@{}}Non-Zeros \\ \ic{nonzero}\end{tabular} \\ \hline
\begin{tabular}[c]{@{}c@{}}Vanilla \\ DTensor\end{tabular} &
  \begin{tabular}[c]{@{}c@{}}550 $\mu$s\\ (100\%)\end{tabular} &
  \begin{tabular}[c]{@{}c@{}}70 $\mu$s\\ (100\%)\end{tabular} &
  \begin{tabular}[c]{@{}c@{}}580 $\mu$s\\ (100\%)\end{tabular} &
  \begin{tabular}[c]{@{}c@{}}133 $\mu$s\\ (100\%)\end{tabular} \\ \hline
\begin{tabular}[c]{@{}c@{}}Rule-\\ Based \\ Bypass\end{tabular} &
  \begin{tabular}[c]{@{}c@{}}550 $\mu$s\\ (100\%)\end{tabular} &
  \begin{tabular}[c]{@{}c@{}}70 $\mu$s\\ (100\%)\end{tabular} &
  \begin{tabular}[c]{@{}c@{}}580 $\mu$s\\ (100\%)\end{tabular} &
  \begin{tabular}[c]{@{}c@{}}45 $\mu$s\\ (33.8\%)\end{tabular} \\ \hline
\begin{tabular}[c]{@{}c@{}}Sharding \\ Prop. \\ Cache\end{tabular} &
  \begin{tabular}[c]{@{}c@{}}68 $\mu$s\\ (12.4\%)\end{tabular} &
  \begin{tabular}[c]{@{}c@{}}65 $\mu$s\\ (92.9\%)\end{tabular} &
  \begin{tabular}[c]{@{}c@{}}63 $\mu$s\\ (10.9\%)\end{tabular} &
  \begin{tabular}[c]{@{}c@{}}98 $\mu$s\\ (73.7\%)\end{tabular} \\ \hline
\begin{tabular}[c]{@{}c@{}}Efficient \\C++ \\ Core\end{tabular} &
  \begin{tabular}[c]{@{}c@{}}35 $\mu$s\\ (4.5\%)\end{tabular} &
  \begin{tabular}[c]{@{}c@{}}38 $\mu$s\\ (38.8\%)\end{tabular} &
  \begin{tabular}[c]{@{}c@{}}38 $\mu$s\\ (5.49\%)\end{tabular} &
  \begin{tabular}[c]{@{}c@{}}38 $\mu$s\\ (28.5\%)\end{tabular} \\ \hline
\begin{tabular}[c]{@{}c@{}}Static \\ Eager \\ Mode \end{tabular} &
  \begin{tabular}[c]{@{}c@{}}0 $\mu$s\\ (0\%)\end{tabular} &
  \begin{tabular}[c]{@{}c@{}}0 $\mu$s\\ (0\%)\end{tabular} &
  \begin{tabular}[c]{@{}c@{}}0 $\mu$s\\ (0\%)\end{tabular} &
  \begin{tabular}[c]{@{}c@{}}0 $\mu$s\\ (0\%)\end{tabular} \\ \hline
\end{tabular}
\end{table}